\documentclass[journal,12pt, draftclsnofoot, onecolumn]{IEEEtran}
\usepackage{graphicx}
\usepackage{amsmath,amssymb,dsfont,bbm,epstopdf,pgfplots,mathtools,enumitem,mathrsfs, bbm, algpseudocode , graphicx} 
\usepackage[normalem]{ulem}
\usepackage{color}
\usepackage{cite}%hyperref}
\usepackage{graphics}
\usepackage{tikz}
\usepackage{pgfplots}
\usepackage{subcaption}
\usetikzlibrary{shapes, arrows, decorations.markings, arrows.meta}
\usetikzlibrary{patterns} % LATEX and plain TEX when using Tik Z
\allowdisplaybreaks
\graphicspath{{.}{./Figures/}} % path to figures
\allowdisplaybreaks[4] % Allows the align environment to wrap over pages.

\usepackage{graphicx}
\usepackage{amsmath,amssymb,dsfont,bbm,epstopdf,pgfplots,mathtools,enumitem,mathrsfs, bbm, algpseudocode , graphicx} 
\usepackage[normalem]{ulem}
\usepackage{color}
\usepackage{cite}%hyperref}
\usepackage{graphics}
\usepackage{tikz}
\usepackage{pgfplots}
\usepackage{subcaption}
\usetikzlibrary{shapes, arrows, decorations.markings, arrows.meta}
\usetikzlibrary{patterns} % LATEX and plain TEX when using Tik Z
\allowdisplaybreaks
\graphicspath{{.}{./Figures/}} % path to figures
\allowdisplaybreaks[4] % Allows the align environment to wrap over pages.

\newcommand{\D}{\text{D}}
\newcommand{\Dt}{\D_{\text {Tx}}}
\newcommand{\Dr}{\D_{\text {Rx}}}
\newcommand{\muT}{{\mu}_\Tx}
\newcommand{\muR}{{\mu}_\Rx}

\renewcommand{\S}{{\sf{S}}}

\renewcommand{\i}{{\iota}}

\newcommand{\U}{\mathsf{U}}
\newcommand{\e}{\mathsf{e}}

\newcommand{\K}{\mathcal K}

\newcommand{\vect}[1]{\boldsymbol{#1}}

\newcommand{\Tx}{\textnormal{Tx}}
\newcommand{\Rx}{\textnormal{Rx}}

\definecolor{ForestGreen}{rgb}{0.0, 0.5, 0.0}
\newcommand{\mw}[1]{{\color{black}#1}}

\renewcommand{\P}{\mathsf{P}}
\renewcommand{\L}{\mathsf{L}}

\newcommand{\tk}{k}

\newcommand{\C}{\mathsf{C}}
\newcommand{\E}[1]{\mathbb{E}\left[#1\right]}

\begin{document}
\title{An Information-Theoretic View of Mixed-Delay Traffic in 5G and 6G}

\author{ \normalsize \IEEEauthorblockN{ Homa Nikbakht$^{1}$, Mich\`ele Wigger$^2$,  Malcolm Egan$^1$,  Shlomo Shamai (Shitz)$^3$,  \\ Jean-Marie Gorce$^1$,  and H.~Vincent Poor$^4$}\\ 
	\IEEEauthorblockA{$^1$Université de Lyon, INRIA, INSA, CITI, EA3720, 69621 Villeurbanne, France \\   $^2$LTCI,   T$\acute{\mbox{e}}$l$\acute{\mbox{e}}$com Paris, Institut polytechnique de Paris, 91120 Palaiseau, France \\ $ ^3$Department of Electrical Engineering, Technion–IIT, Haifa 3200003, Israel \\  $^4$School of Engineering and Applied Science, Princeton University, Princeton, NJ 08544, USA\\ 
		\{homa.nikbakht, malcom.egan,  jean-marie.gorce\}@inria.fr, michele.wigger@telecom-paris.fr,  sshlomo@ee.technion.ac.il, poor@princeton.edu}}
\maketitle

\begin{abstract}
%% Here goes the abstract
Fifth generation mobile communication systems (5G) have to accommodate both Ultra-Reliable Low-Latency Communication (URLLC) and enhanced Mobile Broadband (eMBB) services. While, eMBB applications support high data rates,  URLLC services aim at guaranteeing low-latencies and high-reliabilities.   eMBB and URLLC services are scheduled on the same frequency band, where the different latency requirements of the  communications  render the coexistence  challenging.
In this survey, we review,  from an information theoretic perspective, coding schemes that  simultaneously accommodate  
 URLLC and eMBB transmissions and show that they outperform traditional scheduling approaches. Various communication scenarios are considered, including point-to-point channels, broadcast channels, 
interference networks, cellular models, and cloud radio access networks (C-RANs). The main focus is on the set of rate pairs  that
 can simultaneously be achieved for URLLC and eMBB messages, which well captures the tension 
 between the two types of communications. We also discuss finite-blocklength results where the measure of interest 
 is the set of error probability pairs that can simultaneously be achieved on the two  communication regimes. % for 
 %given message sizes. 
%The focus is to evaluate the simultaneous rate pairs,  degree of freedom (DoF) pairs and error probability pairs of eMBB and URLLC communications over these models.  
\end{abstract}
\section{Introduction}

{\color{black}Modern communication networks serve a range of applications with heterogenous characteristics. Indeed, 5G and proposed 6G wireless mobile cellular networks are expected to serve a diverse set of applications including telephony, video-streaming, online gaming, time-critical control for transportation or remote surgery, or massive machine-type applications for sensor networks in the Internet of Things (IoT) \cite{Tataria201216g}. These applications differ in terms of both reliability and latency requirements. A key example is when \emph{Ultra-Reliable Low-Latency Communication (URLLC)} and \emph{enhanced Mobile Broadband (eMBB)} \cite{Popovski20185g} applications utilize the same time-frequency resource blocks.

	URLLC is designed to ensure 99.99\% reliability at a maximum end-to-end delay of no more than one millisecond \cite{Popovski2019, Ge2019, Shirvanimoghaddam2019, Zhang2019, Bairagi, Xiao2021, Bennis2018, ZZhang2019}. It is thus suited for delay- and mission-critical applications such as remote surgery, control of manufacturing sites, or communication to and from autonomous vehicles. On the other hand, eMBB is most prominently used for video streaming and other applications with less stringent delay tolerances \cite{Popovski20185g}.
	
	 In 5G and proposed 6G systems, URLLC and eMBB users are allocated \textit{network slices}, which correspond to resources within the radio access network. A key challenge is how to design resource allocation and coding schemes given that URLLC and eMBB slices have very different delay requirements. As such, the network must support \textit{mixed delay traffic}. This challenge is further complicated when the radio access network exploits advanced architectures, such as {\emph{cloud radio access networks (C-RANs)} \cite{Checko2014cloud, Peng2015}  (illustrated in Fig.~\ref{fig:CRAN})  or cooperative networks (illustrated in Fig.~\ref{fig:Coop_Net}).
	
	One standard approach is to use smart scheduling and resource allocation algorithms, which interrupt eMBB transmissions to send URLLC data. Various scheduling algorithms have been proposed, which exploit machine learning techniques \cite{Zhang20192, Bairagi2019,Elsayed2019,Alsenwi2021} (including deep learning \cite{Khan2020,Li2020}) and intelligent reflective surfaces \cite{Almekhlafi2021} to improve performance.

\begin{figure}[t!]
\begin{subfigure}{0.5\textwidth}
		\begin{center}
\includegraphics[height=8cm]{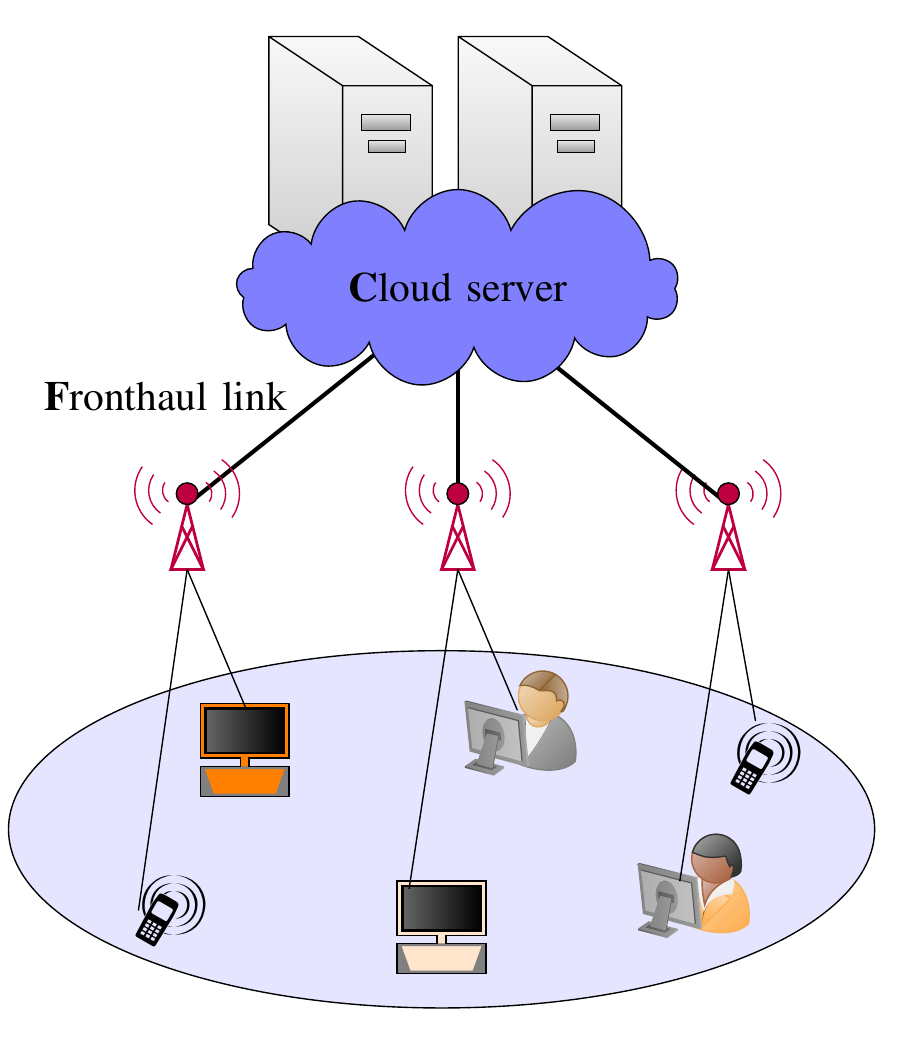}
\end{center}
\caption{C-RAN}
\label{fig:CRAN}
\end{subfigure}
\begin{subfigure}{0.5\textwidth}
\vspace{1.5cm}

\begin{center}
\includegraphics[width=8cm]{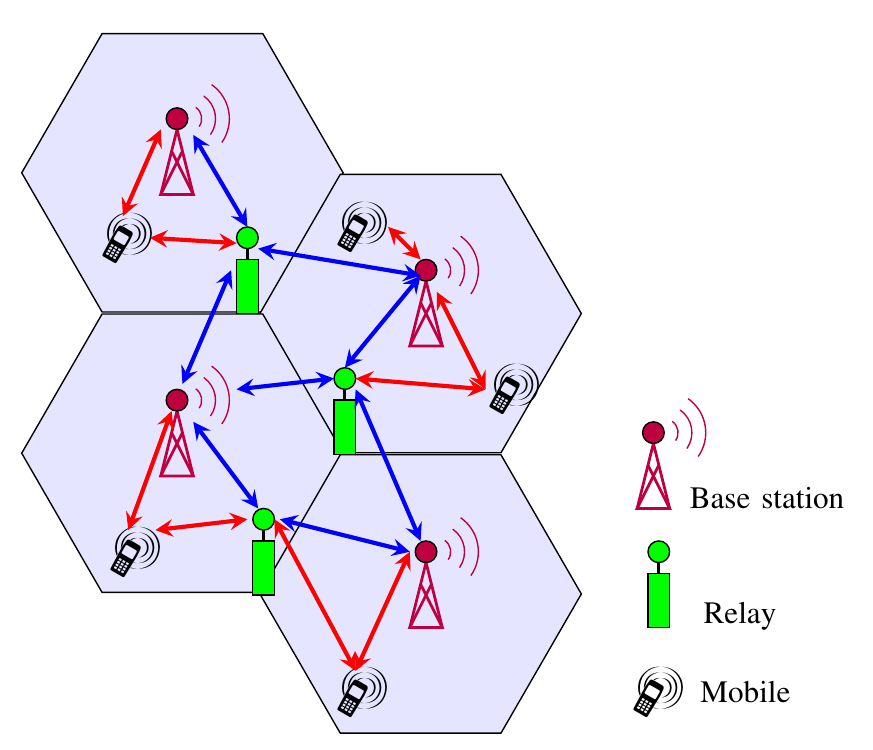}
\caption{Cellular network with cooperation links}
\label{fig:Coop_Net}
\end{center}
\end{subfigure}
\caption{Infrastructures  allowing to mitigate interference.}
\label{fig:infrastructure}
\end{figure}
	
	Nevertheless, an \textit{information theoretic perspective} suggests that performance can be further improved via advanced \textit{joint} coding schemes, which account for the mixed delay requirements of URLLC and eMBB slices. % and utilize the front- and back-haul links available in architectures such as C-RAN) and cooperative networks (illustrated in Fig.~\ref{fig:Coop_Net}). %Important examples of this principle arise in the broadcast channel, relying for example on superposition coding \cite{CoverBC}.  
	Indeed, by introducing joint coding schemes, data from both URLLC and eMBB slices can be simultaneously transmitted in the same resource  block, as illustrated in Fig.~\ref{fig:approaches}. A key issue is that interference is introduced not only by multiple users within the same time-frequency resource, but also from data from each slice transmitted by the \textit{same} user. Nevertheless, as we show in this survey, by using appropriate joint coding techniques, the presence of interference does not necessarily lead to reductions in performance. % By exploiting centralized processing in C-RANs, interference can be beneficial as it creates alternative communication paths from users to the central cloud processor, leading to improved performance. 
	
	In this survey, we overview recent work on joint coding with mixed delay traffic arising from URLLC and eMBB slices in network architectures ranging from point-to-point to C-RANs. %Our focus is primarily on joint coding schemes based on superposition and dirty-paper coding \cite{Costa1983}. As such, we do not consider rate-splitting schemes, which have recently been surveyed in \cite{Mao2022rate} in the context of multiple-access networks. 

\begin{figure}[t!]
 \begin{subfigure}{0.5\textwidth}
 \begin{tikzpicture}[scale=1.4, >=stealth]
 \centering
\draw[thick, ->] (-0.75,3.75)--(4.75,3.75);
\draw[thick] (-0.75,3.65)--(-0.75,3.85);
{\node [draw=none] at (5,3.75) {$ t$};}
\draw[line width= 2mm, blue] (-0.5,3.75)--(1,3.75);
\draw[line width= 2mm, red] (1,3.75)--(2.5,3.75);
\draw[line width= 2mm, blue] (2.5,3.75)--(4,3.75);
\draw[line width= 1mm, <-, red] (1,3.75)--(1,4.2);
\draw[line width= 1mm, ->, blue] (4,3.75)--(4,3.3);
\draw[line width= 1mm, ->, red] (2.5,3.75)--(2.5,3.3);
\draw[line width= 1mm, <-, blue] (-0.5,3.75)--(-0.5,4.2);
\node [draw=none] at (0.2,4.1) {\footnotesize eMBB};
\node [draw=none] at (0.2+1.5,4.1) {\footnotesize{URLLC}};
\node [draw=none] at (0.2+3,4.1) {\footnotesize eMBB};
\end{tikzpicture}
\caption{Scheduling approach}
\end{subfigure}
 \begin{subfigure}{0.5\textwidth}
 \begin{tikzpicture}[scale=1.4, >=stealth]
 \centering
\draw[thick, ->] (-0.75,3.75)--(4.75,3.75);
\draw[thick] (-0.75,3.65)--(-0.75,3.85);
{\node [draw=none] at (5,3.75) {$ t$};}
\draw[line width= 2mm, blue] (-0.5,3.75)--(1,3.75);
\draw[line width= 1mm, blue] (1,3.715)--(2.5,3.715);
\draw[line width= 1mm, red] (1,3.7875)--(2.5,3.7875);
\draw[line width= 2mm, blue] (2.5,3.75)--(4,3.75);
\draw[line width= 1mm, <-, red] (1,3.75)--(1,4.2);
\draw[line width= 1mm, ->, blue] (4,3.75)--(4,3.3);
\draw[line width= 1mm, ->, red] (2.5,3.75)--(2.5,3.3);
\draw[line width= 1mm, <-, blue] (-0.5,3.75)--(-0.5,4.2);

\node [draw=none] at (0.2,4.1) {\footnotesize eMBB};
%\node [draw=none] at (0.2+1.5,4.1+0.5) {\footnotesize{URLLC}};
%\node [draw=none] at (0.2+1.5,4.1+0.3) {\footnotesize{\&}};
\node [draw=none] at (0.2+1.6,4.1) {\footnotesize{eMBB$+$URLLC}};
\node [draw=none] at (0.2+3,4.1) {\footnotesize eMBB};
%\node [draw=none] at (0.2+1.5,4.1+0.3) {URLLC};
\end{tikzpicture}
\caption{Joint coding approach}
\end{subfigure}
\caption{URLLC and eMBB transmissions: (a) Scheduling approach, (b) Joint coding approach. }
\label{fig:approaches}
\end{figure}
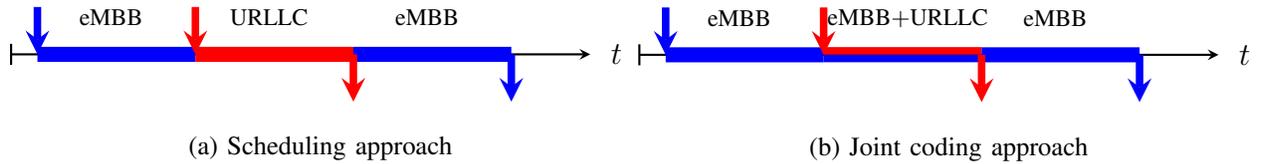
	
\subsection{Related Works}
While in this survey we focus on communication scenarios with different network slices that have different delay constraints, information-theorists}} have also studied related scenarios with other types of heterogeneous communication requirements. The works most closely related to mixed-delay are  \cite{Steinberg2017, Keresztfalvi2019, Keresztfalvi2020}. Specifically,  \cite{Steinberg2017} studies a scenario where two messages are transmitted over a broadcast channel, but only one of them can profit from the cooperation link between the two receivers. The other message has to be decoded directly based on the legitimate receiver's channel outputs, without cooperation from the other receiver.  The motivation in \cite{Steinberg2017} to study such a system was to design a robust communication scheme  where the receivers can reliably decode two messages when the cooperation link between the receivers is present, while still being able to reliably decode  a single message in case the cooperation link fails. A different interpretation, but with the same mathematical model, is to say that one of the messages needs to be decoded immediately without waiting for the cooperation message from the other receiver, while the other  message can tolerate more delay and therefore be decoded also based on the cooperation message. In this sense, the model \cite{Steinberg2017} well suits also a mixed-delay communication scenario  with URLLC and eMBB slices as  considered  in this survey. {\color{black}The work in \cite{Keresztfalvi2019,Keresztfalvi2020} study a scenario with different reliability criteria of two slices, as also characteristic  for URLLC and eMBB slices. In particular,  \cite{Keresztfalvi2019} imposes the constraint that the data from one slice has to be decoded even under an adversarial attack model (the arbitrarily varying channel \cite{AVC, AVCNarayan, AVCLapidoth}) whereas the data from the  other slice only has to be decoded in the likely event that the channel exhibits an expected behavior. %The work  in \cite{Langberg2021} studies a network consisting of multiple transmitters and receivers where each receiver has different decoding deadlines depending on the type of its desired data. 

		\subsection{Related Surveys and Contributions}
	
	\begin{table}\label{table:survey}
		\caption{Related Surveys}
		\begin{center}
			\begin{tabular}{| l | c | r |}
				\hline			
				Survey & Year & Comments \\
				\hline
				\cite{Morais2020when} & 2020 & System level perspective on C-RANs. \\
				\hline
				\cite{Checko2014cloud} & 2014 & System level perspective on C-RANs. \\
				\hline  
				\cite{Zhang2019overview} & 2019 & Overview of network slicing.\\
				\hline
				\cite{Hossain2015cellular} & 2015 & Overview of 5G cellular interference management.\\
				\hline
				\cite{Sun2019application} & 2019 & Machine learning for interference management.\\
				\hline
				\cite{Mao2022rate} & 2022 & Survey on rate-splitting in multiple access networks..\\
				\hline
				\cite{Popovski20185g} & 2018 & Overview of eMBB and URLLC from a communications theory perspective.\\
				\hline
				\cite{Shariatmadari2018fifth} & 2018 & Survey on control channel design.\\
				\hline
				\cite{Vaezi2021cellular} & 2021 & Survey on communication theoretic aspects of 5G.\\
				\hline
				\cite{Makki2020survey} & 2020 & Survey on NOMA.\\
				\hline
				\cite{Vaezi2019interplay} & 2019 & Survey on NOMA.\\
				\hline
				\cite{Dai2018survey} & 2018 & Survey on NOMA.\\
				\hline 
				\cite{Islam2016power} & 2016 & Survey on NOMA.\\
				\hline			
			\end{tabular}
\label{table:survey}
		\end{center}
	\end{table}
	
	{ A number of surveys, summarized in Table~\ref{table:survey}, have recently appeared covering varying aspects of coding, resource allocation, and architecture design in 5G and beyond. The surveys in \cite{Morais2020when,Checko2014cloud,Zhang2019overview} have focused on system level approaches in order to support network slicing, but do not consider aspects related to coding.
	
	On the other hand, the surveys in \cite{Hossain2015cellular,Sun2019application,Popovski20185g,Vaezi2021cellular,Makki2020survey,Vaezi2019interplay,Dai2018survey,Islam2016power,Mao2022rate} focus on communication theoretic aspects of 5G and future 6G systems. In particular,  \cite{Hossain2015cellular,Sun2019application,Popovski20185g,Vaezi2021cellular} consider various resource allocation techniques and  specifically  \cite{Popovski20185g}  treates resource allocation for eMBB and URLLC slices. 
%	The closest surveys to the present paper are \cite{Mao2022rate,Makki2020survey,Vaezi2019interplay,Dai2018survey,Islam2016power}. 
	The surveys in \cite{Makki2020survey,Vaezi2019interplay,Dai2018survey,Islam2016power} focus on non-orthogonal multiple access (NOMA) schemes based on successive interference cancellation for multiple access networks, while  %These schemes exploit , which we also consider; however, the survey is limited to multiple access systems and does not exploit superposition coding. 
	the survey  \cite{Mao2022rate} overviews the benefits of rate-splitting techniques on the multiple-access channel \cite{Mao2022rate}. % is the emphasis on the multiple access channel. In contrast, our survey considers recently proposed architectures such as C-RAN, taking into account the present of front- and back-haul links. 
		
	Despite the importance of joint coding schemes with mixed delay traffic, there has not been a comprehensive survey on this topic. This survey aims to fill this gap by highlighting how joint coding can improve performance of standard scheduling schemes, drawing on fundamental insights from an information theoretic analysis of the network.
	
	The main contributions in this survey are summarized as follows:
	\begin{enumerate}
		\item[(i)] An overview of interference mitigation techniques drawn from information theory, with a focus on superposition coding, dirty paper coding, and coordinated multi point transmission and reception.
		\item[(ii)] A summary of joint coding schemes and recent results on their performance for mixed delay traffic in 
		\begin{enumerate}
			\item[(a)] point-to-point networks;
			\item[(b)] broadcast networks;
			\item[(c)] cooperative networks;
			\item[(d)] C-RANs.
		\end{enumerate}
		\item[(iii)] A discussion of open problems in the design of joint coding schemes for mixed delay traffic. 
	\end{enumerate} 
	
	\subsection{Outline of the Survey}}
	
	This survey article is organized as follows. In Section~\ref{sec:Preliminaries} we review known interference mitigation techniques such as superposition coding, dirty-paper coding, and Coordinated Multi-Point (CoMP) transmission and reception. For a more thorough discussion of these tools, we refer to the original articles or  standard textbooks  \cite{CoverThomas06,ElGamal-Book}. We then continue in Section~\ref{sec:P2P} to discuss integrated transmission of URLLC and eMBB messages on P2P channels with a single transmitter and a single receiver, followed by Section~\ref{sec:BC} that discusses extensions to multi-receiver broadcast channels (BC). 
	Sections~\ref{sec:Network} and \ref{sec:CRAN} consider mixed-delay transmissions over cooperative cellular interference networks and C-RANs.
	The survey is concluded with a summary and outlook section in Section~\ref{sec:Summary}.
	
	\textit{Notation:} Throughout the survey, we abbreviate \emph{transmitter} and \emph{receiver} by \emph{Tx} and \emph{Rx}. For \emph{independent and identically distributed} we use \emph{i.i.d.}. Random variables are  denoted using upper case letters, and realizations thereof by lower case letter, e.g., $X$ and $x$. Random vectors are denoted with uppercase bold symbols.  Fixed constants are often written with sans-serif font, for example $\mathsf{K}, \P, \mathsf{Q}$ or using Greek letters, for example $\rho$ and $\alpha$.  To follow standard notation we however use $n$ to denote the blocklength of transmission. 
For any positive integer $\mathsf{K}$ we use the short-hand notation $[\mathsf{K}]=\{1,\ldots, \mathsf{K}\}$. 
Channels are assumed to be real-valued, extensions to complex channels with independent real and complex components are straightforward. In this sense, $\mathcal{N}(0,\sigma^2)$ denotes the real centered Gaussian distribution of  variance $\sigma^2$. We also use the usual shorthand notation $Y^n=(Y_1,\ldots, Y_n)$.

%	\bibliographystyle{ieeetr}
%	
%	\bibliography{homa_tutorial.bib}
%	
%	
%\end{document}

}

\section{{Mixed Delay Traffic and Interference Mitigation}}\label{sec:Preliminaries}
%Both in P2P and network scenarios, there are situations where communications of different data streams occupy the same time slots and frequency bands, and thus interfere with each other. Managing interference between URLLC and eMBB communications is one of the main topics of this survey paper. As we will see, the main challenge is to appropriately combine existing interference management tools and to adapt them so as to account for the heterogeneous latency requirements, and to mathematically analyze these complex schemes.  In this section, we review some existing interference management techniques which will be used in latter sections.

{\color{black}
\subsection{Coding and Delay}

	The primary goal of a communications network is to reliably send one or more messages $M_i \in \{1,\ldots, \mathsf{M}_i\}$ from one or more Txs to one or more Rxs. To do so, each Tx \emph{encodes} the different messages it wishes to send into a  waveform $x^n_k$, which corresponds to the physical signal sent over the network. In a scenario with \emph{homogeneous delay constraints} i.e., where all messages are sent over the same blocklength $n$,  a Tx $k$ encodes  its  messages $\{M_{i} \colon i \in \mathcal{T}_{k}\}$, where $\mathcal{T}_k$ collects the indices of all messages sent by Tx $k$,  into the codeword $x_k^n(\{M_i \colon i \in \mathcal{T}_k\})$ using a joint encoding function $f_k \colon \{1,\ldots, \mathsf{M}_{i_k} \} \times \cdots \times   \{1,\ldots, \mathsf{M}_{i_{k+1}-1}\} \to \mathbb{R}^{n}$. After receiving the corresponding output symbols $y_k^n$,  Rx $k$ produces a guess $\{\hat{M}_i \colon i \in \mathcal{R}_k\}$ for each of its desired messages  $\{M_i \colon i \in \mathcal{R}_k\}$, where $\mathcal{R}_k$ collects the indices of the messages intended for Rx $k$,  by applying a decoding function $g_k\colon \mathbb{R}^n \to  \{1,\ldots, \mathsf{M}_{i_k} \} \times \cdots \times   \{1,\ldots, \mathsf{M}_{i_{k+1}-1}\}$  to its observed outputs $y_k^n$.
	
	 A more complicated scenario typically arises in  the  mixed-delay scenarios we consider in this survey,  because the various messages are created at  different  times and have different decoding delays. In this case, each message is assigned a creation time $a_i$ and a latest possible decoding time $d_i$. As a consequence, a Tx $k$  produces its inputs $x_k^n$ using per-symbol encoding functions $\{f_{k,t}\}_t$, where at time-$t$  the function $f_{k,t}$ maps all its previously created messages  to an input symbol: 
	 \begin{equation}
	 x_{k,t}=f_{k,t} ( \{ M_{i} \colon i \in \mathcal{T}_k,  a_i \leq t\}).
	 \end{equation} Rx $k$ decodes any of its intended messages  $\{M_i\colon i \in \mathcal{R}_k\}$ by applying the decoding  function $g_i$ to the first $d_i$ channel outputs $y_k^{d_i}$. The decoding function that produces the message guess $\hat{M}_i$ is thus of the form $g_i\colon \mathbb{R}^{d_i} \to \{1,\ldots, \mathsf{M}_i\}$.

	%Assume that this Tx sends the signal $x^{n_i}$ over a channel denoted by $P_{y^{n_i}|x^{n_i}}$. If a Rx wishes to decode $m_i$, it uses the output sequence $y^{n_i}$ and estimates $m_i$ by 

%\begin{equation}
%\hat m_i = g_i(y^{n_i}),
%\end{equation}
%where the function $g_i:  \mathbb{R}^{n_i} \rightarrow \{1,\ldots,M_i\} $ is called the \textit{decoding function}. 
%
%	
	%For our purposes, each codeword is a vector $x^{n_i} \in \mathbb{R}^{n_i}$. If a transmitter wishes to send message $m_i$, then it produces the unique codeword $x^{n_i}(m_i)$, indexed by the message $m_i$. The function $f_i: \{1,\ldots,M_i\} \rightarrow \mathbb{R}^{n_i}$ is called the \textit{encoding function}. 
	
	Associated with the described mixed-delay encodings  are two key parameters:
	\begin{enumerate}
		\item[(i)] the length of each codeword, $n_i=d_i-a_i$;
		\item[(ii)] and the \textit{rate}, defined by 
		\begin{align}
		R_i = \frac{\log \mathsf{M}_i}{n_i}.
		\end{align}
		%where $M_i$ is implicitly a function of $n_i$. 
	\end{enumerate} 
	
	In multi-hop scenarios such as experienced in C-RANs or cooperative networks, transmission delay not only depends on the blocklength of communication, but also on the  delay introduced from the communication over the addition hops. For example, in the uplink of C-RANs, the total delay  experienced for the transmission of a messages  is formed by:
	\begin{itemize}
	\item the communication delay over the network from the mobile users to the BSs; 
	\item the processing time of the compression at the BSs as well as  the communication delay over the fronthaul links to the cloud processor;
	\item the decoding processing time  at the cloud processor. 
	\end{itemize}
In mixed-delay networks, the additional delay introduced by the compression at the BSs and the  fronthaul communication might exceed the latest allowed decoding time $d_i$ for  certain messages $M_i$, which thus have to be directly decoded at the BSs. A similar situation is also encountered  in the downlink of C-RANs, where URLLC messages should directly be encoded at the BSs and not at the cloud processor so as to avoid the  delay introduced by the additional  communication hop over the fronthaul link.  In the same way, URLLC messages transmitted   in cooperative interference networks cannot support the additional communication hops required to establish cooperation between Txs or Rxs. In these networks, the cooperative communication at the Tx side thus can only depend on eMBB messages and the cooperative communication at the Rx side can only serve decoding of eMBB messages. We will provide a more detailed model for the encoding and decoding of mixed-delay messages in Section~\ref{sec:Network} ahead. A main assumption in our model will be  that the communication over the interference network is sufficiently short so  that also URLLC messages can tolerate the introduced delay.

	As we will overview in this survey, mixed delay constraints require careful design of the joint  coding schemes, in particular to  mitigate  the interference caused by the different  slices. In the remainder of this section, we summarize key information-theoretic interference mitigation  techniques. }% in order to construct codes that are tailored to mixed delay constraints.}

\subsection{Superposition Coding} 

\begin{figure}[t!]
\begin{center}
\includegraphics[width=.66\textwidth]{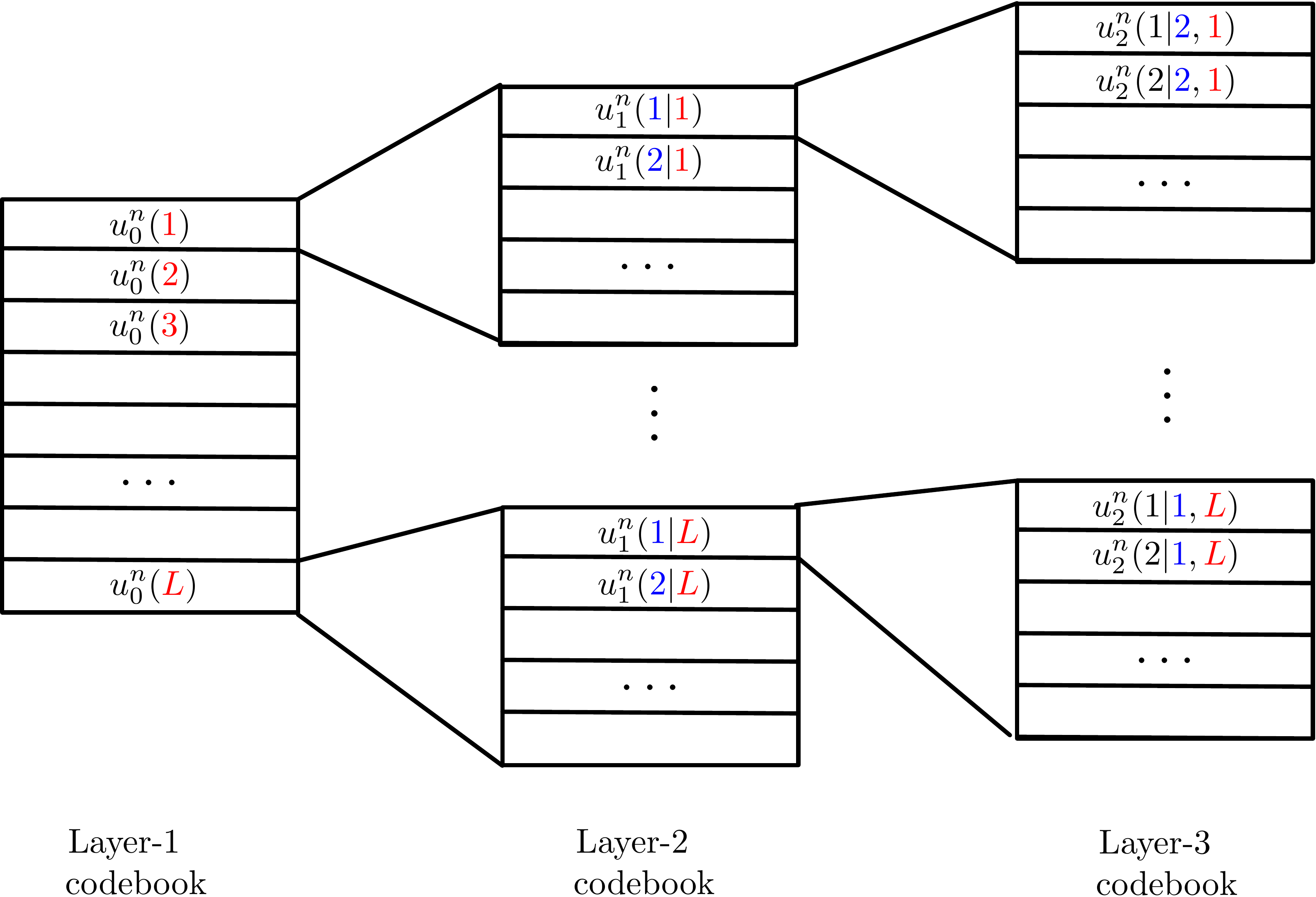}
\caption{Superposition code with 3 layers.} 
\label{fig:superpos}
\end{center}
\end{figure}

Superposition coding\cite{CoverBC, ElGamal-Book} was first proposed in the context of broadcast communication. It can be used to send multiple messages to one or more receivers. The main technical feature is that the different messages are encoded into the different layers of a so called superposition code, illustrated in  Figure~\ref{fig:superpos} for a code example with three layers. The entries of the layer-1 code are  drawn i.i.d. according to a chosen distribution $P_{U_0}$. \emph{For each layer-1 codeword $u_0^{(n)}(\ell)$ a new layer-2 codebook is chosen.} The entries of the layer-2 codewords are drawn independent of each other and the $i$-th entry follows a conditional distribution $P_{U_1|U_0}$ given the $i$-th entry of   codeword $u_0^{(n)}(\ell)$. For each layer-2 codeword, a new layer-3 codebook is chosen. Entries of this codebook are again independently of each other and drawn according to a conditional distribution $P_{U_2|U_1}$ given the entries in the corresponding layer-2 codeword. 

 In a superposition code, each message is not only protected by its corresponding layer, but also by all underlying layers. Given the structure of the code, any receiver that is interested in decoding a given layer also has to decode \emph{all previous layers}, typically in a joint manner.  Upper layers are not decoded and cause additional disturbance (noise) on the decoding of lower layers.
 
Given the described decoding order, in the context of mixed-delay traffic it is  possible to send URLLC data on lower layers and eMBB  on upper layers but not the other way around, because URLLC messages have to be decoded first, prior to decoding  eMBB messages. 

A simpler alternative to superposition coding is to  encode each message using an independent codebook and to combine the chosen codewords, by means of a predefined mapping,   to form the  sequence of channel inputs.  In particular, for Gaussian channels, messages are encoded into Gaussian codewords and the sum of these codewords is transmitted over the channel. The advantage of this method is that no layering-order of the codewords has to be established a priori. This is particularly convenient in fading channels where the exact channel statistics are not known at time of encoding, and the receiver can decide on the layers  to decode after having estimated the realization of the channel. In this sense, the simpler alternative can allow for  increased \emph{expected rates} over slowly fading channels where the channel variations are limited over the duration of a single codeword. As we shall see, this approach is also highly beneficial for joint transmission of URLLC and eMBB messages where the fading is almost constant over the duration of an URLLC communication but varies significantly during the transmission of eMBB messages. 

\subsection{Dirty-Paper Coding (DPC)} 
 If interference is known at a Tx before communication starts, the Tx can mitigate this interference through \emph{Dirty Paper Coding} \cite{Costa1983, Erez2005, Cohen2002} and achieve the full capacity of the channel without interference. 
 To illustrate,  consider the Gaussian interference channel %\eqref{eq:channel} with $\L_t = \L_r = 1$. For each $k \in \{1, \ldots, K\}$, fix  $h_{k,k} = 1$ and $h_{k,\tilde k} < 1$ and assume that $|\Itx| = 1$. The observed signal by Rx~$k$ is thus
	\begin{equation}\label{eq:channel2}
	Y^n = X^n + W^n + Z^n,
	\end{equation}
	where  $Z^n$ is an i.i.d. standard Gaussian noise sequence and  $W^n$ is memoryless interference sequence, with each component zero-mean Gaussian of power $\mathsf{Q}$. 	If $W^n$ is unknown to both the Tx and the Rx, the interference simply acts as additional noise, and the capacity of the channel  equals $\frac{1}{2}\log(1 + \frac{\P}{1+ \mathsf{Q}})$. If the Rx knows $W^n$, then it can subtract this interference from the outputs and as a consequence the capacity of the channel is the same as without interference, i.e.,  $\frac{1}{2} \log(1 + \P)$. Costa \cite{Costa1983} showed that when $W^n$ is unknown to the Rx but known to the Tx even before the communication starts, then the  capacity of the channel is also equal to the interference-free capacity $\frac{1}{2} \log(1 + \P)$. The coding scheme achieving this performance was termed dirty-paper coding (DPC) and is described in the following. (For an analysis see \cite{Costa1983, ElGamal-Book}.) 

Define the parameter $\alpha: = \frac{\P}{1+\mathsf{Q}}$ and the random variable $U= X+\alpha W$ where $W\sim\mathcal{N}(0,\mathsf{Q})$ and  $X\sim\mathcal{N}(0,\P)$ independent of each other. It can be  verified that 
\begin{IEEEeqnarray}{rCl}
I(U;Y) & = & \frac{1}{2} \log \left(\frac{ (\P+\mathsf{Q}+1)(\P+\alpha^2 \mathsf{Q})}{\P \mathsf{Q}(1-\alpha)^2 +(\P+\alpha^2\mathsf{Q})}\right)\\
I(U;W) & = & \frac{1}{2} \log \left( \frac{ \P+ \alpha^2 \mathsf{Q}}{\P}\right)\\
I(U;Y)-I(U;W) & = &\frac{1}{2} \log \left( \frac{\P (\P+\mathsf{Q}+1)}{\P \mathsf{Q}(1-\alpha)^2 + (\P+\alpha^2\mathsf{Q})}\right)= \frac{1}{2} \log (1+ \P)\mathsf{Q}
\end{IEEEeqnarray}

Fix $\epsilon > 0$ arbitrary small. For each $m\in\big[2^{n(I(U;Y)-I(U;W) - \epsilon)}\big]$ generate a bin with $2^{n(I(U;W)+\epsilon/2)}$ codewords $\big\{U^n(j,m)\colon j\in\big[2^{n(I(U;W)+\epsilon/2)}\big]\big\}$ by picking each component of each codeword  i.i.d. according to  $\mathcal N \left (0, \P + \alpha^2 \mathsf{Q}\right)$.  Reveal the codebook consisting of all $2^{n(I(U;Y)-I(U;W) - \epsilon)}$ bins  to the Rx and the Tx.
	
\textit{Encoding:}	To encode a message $M=m$, the encoder looks for a codeword $U^n(j,m)$ in bin $m$ that is jointly typical \cite{ElGamal-Book} (i.e., has approximately the correct joint empirical distribution) with the interference sequence $W^n$. The Tx then forms  $X^n =  U^n(j^*,m) - \alpha  W^n$, where $j^*$ indicates the chosen index, and send  this  sequence $X^n$ over the channel.  Note that the Tx  declares an error if no codeword $U^n(j,m)$ in bin $m$ is jointly typical with  the interference $W^n$.
	
\textit{Decoding:} After observing the sequence $Y^n$, the Rx looks for a codeword $U^n(j,m)$ that is jointly typical with $Y^n$. If a single such codeword exists, the Rx declares $\hat{M}=m$, otherwise it declares an error. 

It can be shown that  with probability tending to 1 as the blocklength $n\to \infty$, the only codeword that is jointly typical with $Y^n$ is codeword $U^n(j^*,m)$ which was selected at the transmitter. The Rx thus not only recovers the correct message $\hat{M}=M$ with high probability, but can also reconstruct the transmitted codeword $U^n(j^*,m)$ with high probability. 

\subsection{Coordinated Multi Point (CoMP)} \label{sec:CoMP}
\emph{Coordinated multi-point (CoMP)} refers to a wide set of techniques that enable either  a set of distributed Txs to jointly encode messages  or a set of distributed Rxs to jointly  decode messages. We will be particularly interested in scenarios where CoMP is facilitated through cooperative communication over dedicated links between transmitters or between receivers. 

\mw{In the case of \emph{CoMP transmission} \cite{Yu2013, Chen2019, HomaSPAWC2019,  Levy2009}, we consider a set of distributed Txs, each having one message to send, and with cooperation links between neighbouring Txs. Before communicating to the Rxs, all Txs convey their messages to a dedicated Tx,  \mw{called} the \emph{master Tx}, which then jointly designs input signals for all Txs (also exploiting its available state-information) and conveys \mw{rate-distortion compressed (lossy) versions of these signals to each  of the Txs. The Txs  \mw{reconstruct the compressed signals and send these signals over the channel to the Rxs}. If cooperation links are of sufficiently high rates, then \mw{the loss of the  compression can be maintained at noise-level and} does not decrease the \emph{degrees of freedom (DoF)}, i.e., the factor in front of the logarithmic expansion of the asymptotic high-SNR sum-capacity. In this  case, the interference channel is \mw{intuitively} transformed into a multi-antenna single-Tx  BC and the DoF is given by the minimum number of Tx and Rx antennas.

In case of \emph{CoMP reception} \cite{Zhou2013, Simeone2009, Simeone2012, Egan2013}, consider a set of distributed Rxs, each wishing to decode one message, and with cooperation links between neighbouring Rxs. Each Rx \mw{applies a lossy compression algorithm to} its observed output signal and describes the compressed signal over the cooperation links to a dedicated Rx, \mw{called} \emph{master Rx}. This master Rx \mw{reconstructs all the compressed signals and jointly decodes all the messages, which it then sends  to their intended Rxs over the cooperation links.} If cooperation links are of sufficiently large  rates, then \mw{lossy compression of the receive signals can be performed so that the loss is maintained} at noise-level, which again has no influence on the DoF of the channel, and corresponds to the DoF of a single-Rx multi-access channel which equals the minimum number of Tx and Rx antennas. 

CoMP transmission and reception can only be used to encode and decode eMBB messages, because the communication over the cooperation links induces significant delay. The delay is in fact given by  twice the number of hops required on the cooperation links to reach the master Tx/Rx from any other Tx/Rx in the network times the communication duration on a single cooperation hop. Since in practical networks also eMBB communication is delay-limited, CoMP transmission and reception can be performed only on small subsets of Txs and Rxs, where the size depends on the maximum allowed number of communication hops. }}

\section{Point-to-Point Communications}\label{sec:P2P}
\subsection{Introduction}
This section focuses on  P2P channels with a single Tx and a single Rx.  Subsection~\ref{sec:superpos} reviews the superposition coding approach  over fading channels in  \cite{shlomo2012ISIT, Cohen2012}. This approach manages to send URLLC messages over single coherence blocks of the fading channel without suffering from a degradation due to the lack of state knowledge at the Tx,  and  simultaneously also sends eMBB messages over multiple coherence blocks, thus exploiting the ergodic behaviour of the channel. This approach is also known as  \emph{broadcast approach} and has been studied for a wide  field of applications, see the recent survey paper \cite{Tajer2021}.  

Subsection~\ref{sec:consecutive} summarizes the results in \cite{WCNC2021long}, \cite{ISIT2022long}, which analyze a similar superposition coding approach but for Gaussian channels and in the finite-blocklength regime. The analysis is based on the concept of “parallel channels” introduced in \cite{Erseghe2016}. % which distinguishes the transmission intervals  with and without URLLC. 

\subsection{The Broadcast Approach over Fading Channels without Transmitter Channel State Information}\label{sec:superpos}
This section is based on the results in \cite{shlomo2012ISIT, Cohen2012, Tajer2021}. Consider a P2P channel where a single Tx wishes to send both URLLC and eMBB messages over a fading channel to a single Rx. Latency requirements impose that transmission of URLLC messages spans only a single coherence time of the fading channel, but transmission of eMBB can  span multiple  coherence blocks and thus  profit from channel diversity. In a single-antenna setup, a simple channel model capturing these constraints is as follows:
\begin{equation}
Y_{b,t} = \sqrt{S_b} \cdot X_{b,t} +Z_{b,t}, \quad b=1,\ldots, \mathsf{B}, \quad t=1,\ldots, \mathsf{T},
\end{equation} 
where $\mathsf{B}$ denotes the number of blocks, $\mathsf{T}$ the channel coherence time, $\{S_b\}$ describe the fading power in the various  coherence blocks and are assumed i.i.d. with probability distribution function (pdf) $f_S$ and variance $1$, and $\{Z_{b,t}\}$ is a sequence of i.i.d. standard Gaussian noises. The fading power is assumed to be perfectly known at the Rx (e.g., by transmitting pilot signals at the beginning of each block based on which the Rx can estimate the fading power), but not at the Tx. Since each URLLC message can be transmitted only over a single coherence block, the channel inputs are formed as 
\begin{equation}
X_{b,t} = f_{b,t} \left(M^{(\U)}_b, M^{(\e)}\right),  \quad b=1,\ldots,  \mathsf{B}, \quad t=1,\ldots, \mathsf{T},
\end{equation}
for some appropriate encoding functions $\{f_{b,t}\}$ satisfying the power constraint 
\begin{equation}
\frac{1}{ \mathsf{B} \mathsf{T}} \sum_{b=1}^\mathsf{B} \sum_{t=1}^{\mathsf{T}} | X_{b,t} |^2  \leq \P,
\end{equation}
and where $M^{(\U)}_b$ indicates the URLLC message sent in block $b$ and $M^{(\e)}$ the single eMBB message sent over the entire $B$ blocks. 

In the following, messages are assumed independent of each other and uniform over message sets $\mathcal{M}_{\U}$ and $\mathcal{M}_{\e}$. In this subsection, $\mathcal{M}_{\U}=\big[2^{\mathsf{T} R_{\U}}\big]$ and $\mathcal{M}_{\e}=\big[2^{\mathsf{T}\mathsf{B}  R_{\e}}\big]$, where $R_{\U}$ and $R_{\e}$ denote the URLLC and eMBB rates of transmission.

After each block $b$, the Rx decodes the URLLC message $M^{(\U)}_b$ sent in this block:
\begin{equation}
\hat{M}^{(\U)}_b = g_b^{(\U)}( Y_{b,1}, \ldots, Y_{b,\mathsf{T}}), \quad b=1,\ldots, \mathsf{B},
\end{equation}
and at the end of the entire communication it also decodes the eMBB message:
\begin{equation}
\hat{M}^{(\e)} = g^{(\e)}( Y_{1,1}, \ldots, Y_{\mathsf{B},\mathsf{T}}),
\end{equation}
for   decoding functions $\{g_{b}^{(\U)}\}$  and $g^{(\e)}$ on appropriate domains. 

In \cite{shlomo2012ISIT,Cohen2012}, the authors propose to encode the two message streams using  simplified superposition coding where both streams are encoded into independent Gaussian codewords, which are then added up for transmission. More precisely, the URLLC message  \emph{in each block} is encoded into multiple layers so that the Rx can decode as many layers as the actual instantaneous fading power $S_b$ permits. (This implies also that depending on the fading realization, certain URLLC messages are not decoded and in practice  have to be retransmitted in the next block.) To allow for  closed-form expressions, an infinite layering approach is employed with layers that are of infinitesimally small power.  %The eMBB message is decoded only after receiving $B$ blocks, and  by the weak law of large numbers, for large blocks $B$ the average fading realization is close to its means with high probability and thus the transmission of the eMBB message can well be adapted to this average channel quality. 

The Tx allocates total power $\beta \P$ to the transmission of the URLLC messages and power  $(1-\beta)\P$ to transmit the eMBB messages. The power distribution to the different URLLC layers is described by a power density $\rho(\cdot)$ satisfying  $\int_{u}  \rho(u) du= \beta \P$, where $\rho(s')$ indicates the (infinitesimely small) power that is assigned to a given layer that is decoded whenever the fading $S_b\geq s'$.  The interference power stemming from  non-decoded URLLC messages under state $S_b=s$ is then given by $s\cdot I(s)$ where
\begin{equation}
I(s):= \int_{u=s}^\infty \rho(u) \mathsf{d} u.
\end{equation}
Since URLLC messages are decoded after each block, and eMBB messages only at the end of the last block $B$,  decoding of URLLC messages not only suffers from the interference of non-decoded URLLC messages, but also from the interference of eMBB messages. The power of this latter interference is equal to $(1-\beta)\P$  independent of the block (since the Tx has no knowledge about $\{S_b\}$ it cannot adapt the power). 

To decode the eMBB message at  the end of the last block $B$, the Rx first  subtracts the contributions of  the codewords  corresponding to the decoded URLLC messages and then decodes the eMBB message based on this difference while accounting for the interference power created by all non-decoded URLLC messages, which  in block $b$ is given by $I(S_b)$. 

A careful  analysis of the infinite-layering approach, see \cite{Cohen2012}, reveals that the expected  rate of the reliably decoded messages (i.e., messages decoded with error probability tending to $0$ as the blocklength $T\to \infty$) can be as high as 
\begin{equation}\label{eq:Ru}
R^{(\U)}=\int_{u=0}^\infty (1- F_S(u)) \frac{ u \rho(u)}{1+u( I(u)+(1-\beta)\P)} \mathsf{d}  u,
\end{equation} 
where $F_S(\cdot)$ denotes the cumulative distribution function (cdf) associated with the pdf $f_S(\cdot)$.  In the denominator of \eqref{eq:Ru}, the term $u(I(u)+(1-\beta)\P)$ indicates the interference power experienced  during the decoding of URLLC messages stemming from the eMBB transmission and the non-decoded URLLC messages.

For a sufficiently large number of blocks $\mathsf{B}$, the following rate is achievable for the eMBB messages:
\begin{equation}\label{eq:Re}
R^{(\e)}=\int_{u=0}^\infty f_S(u) \log \left( 1+ \frac{(1-\beta)\P u}{1+u I(u)}\right) \mathsf{d} u. 
\end{equation}
Here we find $uI(u)$ in the denominator which describes the interference power of the non-decoded URLLC messages. 

Equations \eqref{eq:Ru} and \eqref{eq:Re}  thus determine the maximum achievable (expected) sum-rate $R^{(\U)}+R^{(\e)}$  in function of the  URLLC interference power $I(u)$ (notice that $\rho(u) = -\frac{\mathsf{d}}{\mathsf{d}u} I(u)$), which is a design parameter of the scheme and can be thus be optimized.  It is shown in  \cite{Cohen2012} that the optimal interference power function $I(s)$ among all continuously differentiable functions satisfying the boundary conditions $I(0)=\beta \P$ and $I(\infty)=0$ is given by:
\begin{equation}\label{eq:Istar}
I^*(s) = \frac{1}{2} \left( \frac{-b(s)+ \sqrt{b^2(s)- 4 a(s)c(s)}}{2 a(s)} \right) ,
\end{equation}
for $a(s)=s f_S(s)$, $b(s)=2(1-\beta)\P f_S(s) s^2- (1-F_s(s))$, and  $c(s)=(1-\beta)^2 \P^2 f_S(s) s^3$. 

\begin{figure}[t!]
\vspace{-1.5cm}
\begin{center}
\includegraphics[height=12cm]{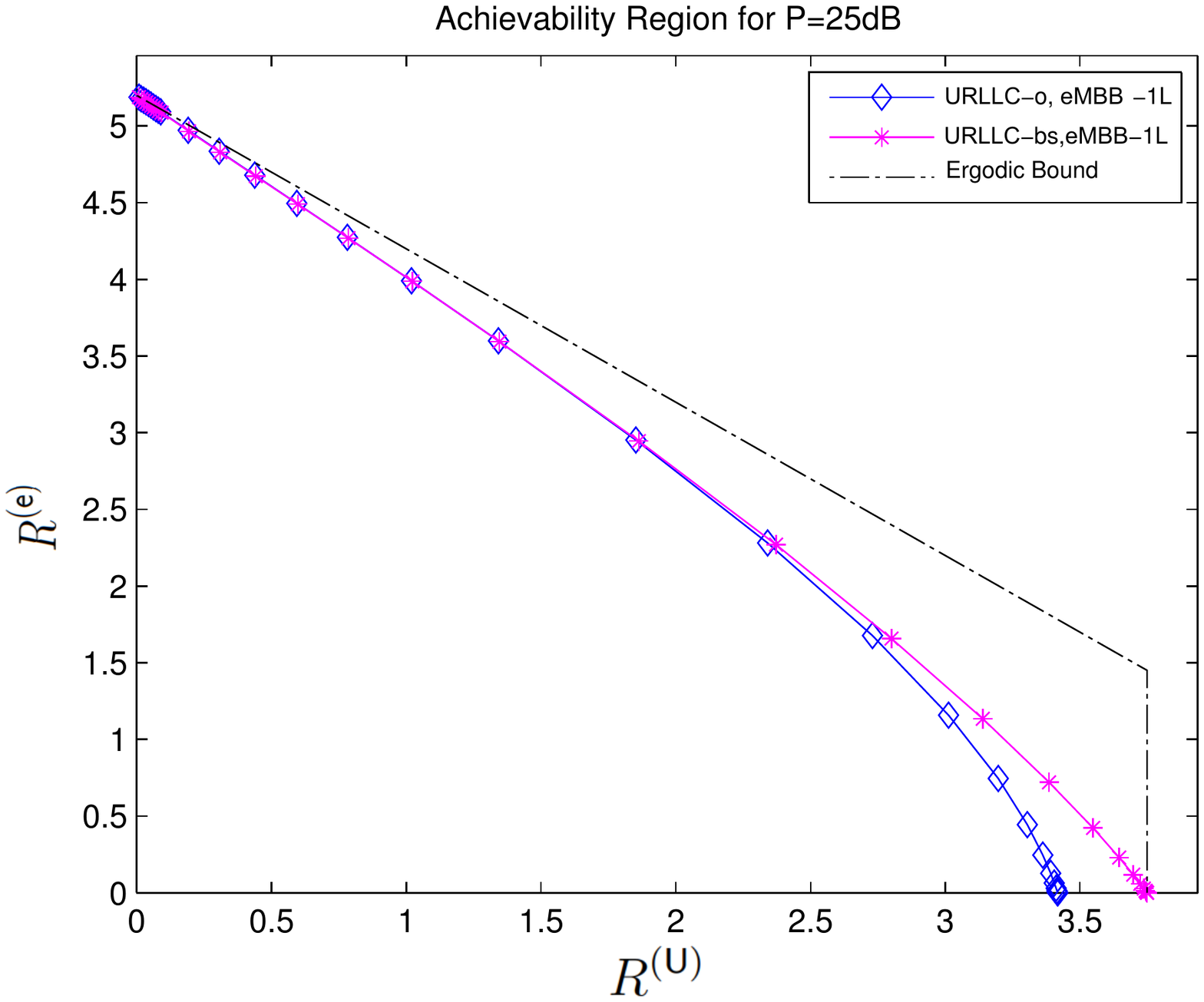}
\end{center}
\vspace{-2.5cm}
\caption{The figure illustrates the set of rate pairs $(R^{(\U)}, R^{(\e)})$ in function of the power allocation parameter $\beta$ and of the interference powers in \eqref{eq:Istar} and \eqref{eq:Io}, respectively.} 
\label{fig:BCapproach}
\end{figure}

Figure~\ref{fig:BCapproach} compares the  sum-rate achieved for this optimal interference power $I^*(s)$ for different power allocation parameters $\beta$ to a simple outage-based approach where $\rho(s)$ is chosen as a  dirac-function at threshold $s_{\textnormal{th}}$, i.e., when  the interference power is  given by the step function
\begin{equation}\label{eq:Io}
I_{\textnormal{o}}(s) = \mathbbm{1}\{s < s_{\textnormal{th}}\}. 
\end{equation}
Here, the  optimal value for the threshold $s_{\textnormal{th}}$ can be derived analytically and is given by the solution to the following equation
\begin{equation}
f_S(s_{\textnormal{th}})\log (1+\beta \P s_{\textnormal{th}}) = (1- F_S(s_{\textnormal{th}})) \frac{ \beta \P}{ (1+\P s_{\textnormal{th}})(1+(1-\beta)\P s_{\textnormal{th}})}.
\end{equation}
From Figure~\ref{fig:BCapproach}, we observe that for small URLLC rates $R^{(\U)}$,  the  penalty in eMBB rates $R^{(\e)}$  is small when using the suboptimal power allocation corresponding to $I_{\textnormal{o}}(s)$ instead of  the optimal  allocation  corresponding to $I^*(s)$. For larger URLLC rates, this penalty increases. 

We further observe that the maximum sum-rate achieved by both power allocations decreases with increasing URLLC rates. The sum-rate for both approaches is more than $5$ when $R^{(\U)}=0$. For  $R^{(\U)}\geq 3.5$  it is  around $4$ under the optimal power allocation  and even vanishes completely under outage power allocation leading to \eqref{eq:Io}. In this high-URLLC-rate regime, the gap to the outer bound is also significant, leaving open the possibility of finding better coding schemes.

%So far, we assumed a Rx that after each block decodes the number of URLLC layers corresponding to the instantaneous fading power $S_b$, and at the end of the last block $B$ it removes all  decoded URLLC layers and finally decodes the eMBB message. 

The described broadcast approach can further be improved by applying a multi-layering approach also for the transmitted eMBB message. In this approach,  different eMBB layers are decoded successively, and after each eMBB decoding step, the Tx decodes further URLLC layers so as to remove their interference for the decoding of subsequent eMBB layers. This additional decoding of URLLC messages at the end of block $\mathsf{B}$ cannot be used to improve performance of the URLLC communication, because the admissible delay is exceeded. However, it allows to  improve decoding performance of eMBB messages. 

{Another way to improve this broadcast approach is to combine it with adaptive \mw{causal} network coding. For example, the work in \cite{Cohen2022} proposes a novel layering scheme consisting of a base layer and an enhancement layer for data streaming under mixed-delay constraints. The base layer contains URLLC data and the enhancement layer  contains eMBB data. In the proposed scheme, the \mw{base layer is  encoded  using a broadcast approach, which allows the Rx to  decode the base layer (i.e., URLLC data)} with minimum delay required. \mw{The enhancement layer is encoded using a priori and posteriori forward error correction  so as to  be able to control the throughput-delay trade-off of this communication.     }}

%Some of the ideas in this section were extended to multi-Rx broadcast channels (BC), see \cite{Lin2021}. These extensions are discussed in Section~\ref{sec:BC}.

\subsection{Finite Block-Length Analysis over Gaussian Channels}\label{sec:consecutive}
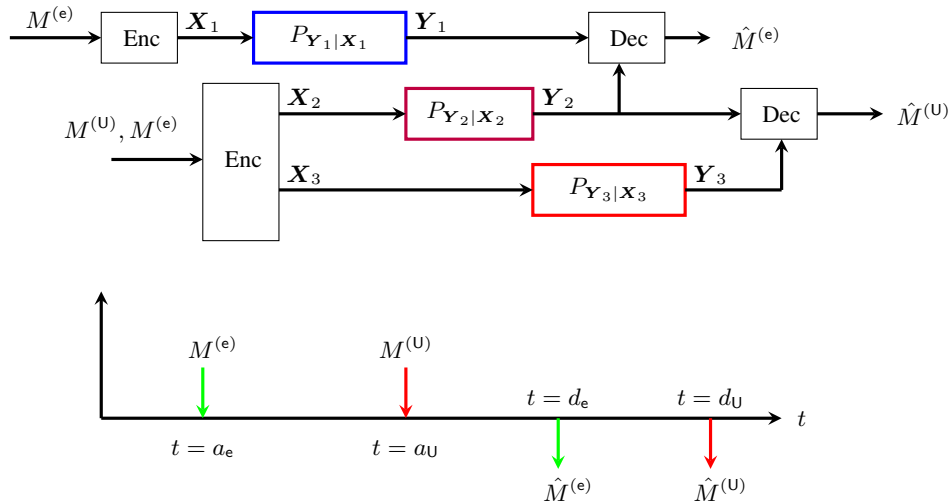
\begin{figure}[t]
%\footnotesize
  \centering
  
\begin{tikzpicture}[scale=1.35, >=stealth]
\centering
\footnotesize
\tikzstyle{every node}=[draw,shape=circle, node distance=0.5cm];
% \foreach \i in {0,...,5} {
% \draw [fill= blue](0 + 0.25*\i,0+1)--(0.25+0.25*\i,0+1)--(0.25+0.25*\i,0.25+1)--(0+0.25*\i,0.25+1)--(0+0.25*\i,0+1);
%}
%\foreach \i in {6,...,10} {
% \draw [fill= purple](0 + 0.25*\i,0+1)--(0.25+0.25*\i,0+1)--(0.25+0.25*\i,0.25+1)--(0+0.25*\i,0.25+1)--(0+0.25*\i,0+1);
%}
%\foreach \i in {11,...,16} {
% \draw [fill= red](0 + 0.25*\i,0+1)--(0.25+0.25*\i,0+1)--(0.25+0.25*\i,0.25+1)--(0+0.25*\i,0.25+1)--(0+0.25*\i,0+1);
%}
%\draw [ line width = 0.5mm] (0, 0.6+1)--(4.25, 0.6+1);
%\draw [ line width = 0.5mm] (0, 0.5+1)--(0, 0.7+1);
%\draw [ line width = 0.5mm] (0+1.5, 0.5+1)--(0+1.5, 0.7+1);
%\draw [ line width = 0.5mm] (0+2.75, 0.5+1)--(0+2.75, 0.7+1);
%\draw [ line width = 0.5mm] (0+4.25, 0.5+1)--(0+4.25, 0.7+1);
%\node[draw =none] (s2) at (0.75,0.75+1) {$n_1$};
%\node[draw =none] (s2) at (2.2,0.75+1) {$n_2$};
%\node[draw =none] (s2) at (3.5,0.75+1) {$n_3$};
%\node[draw =none] (s2) at (1.25+0.2,-0.2+0.8) {$n_p$};
%\draw [ very thick, green , ->] (1.5,-0.2+0.8+0.1)--(1.5,-0.2+0.8+0.4) ; 
%\node[draw =none] (s2) at (1.25+0.2+1.25,-0.2+0.8) {$N_{d_{\e}}$};
%\draw [ very thick, green , ->] (1.5+1.25,-0.2+0.8+0.1)--(1.5+1.25,-0.2+0.8+0.4) ; 
%\node[draw =none] (s2) at (1.25+0.2+1.25+1.5,-0.2+0.8) {$N_{d_{\U}}$};
%\draw [ very thick, green , ->] (1.5+1.25+1.5,-0.2+0.8+0.1)--(1.5+1.25+1.5,-0.2+0.8+0.4) ; 
%
\draw (-1.5+1,-2.75) rectangle (-0.75+1,-1.2);
\draw (-1.5,-1) rectangle (-0.75,-0.5);
\draw[very thick, blue] (0,-1) rectangle (1.5,-0.5);
\draw[very thick, purple] (1.5,-1.75) rectangle (2.75,-1.25);
\draw[very thick, red] (2.75,-2.5) rectangle (4.25,-2);
\draw [ very thick, ->] (-0.75, -0.75)--(0, -0.75);
\draw [ very thick, ->] (-0.75+1, -1.5)--(1.5, -1.5);
\draw [ very thick, ->] (-0.75+1, -2.25)--(2.75, -2.25);
\draw [ very thick, ->] (-2.4+1, -1.75-0.2)--(-1.5+1, -1.75-0.2);
\node[draw =none] at (-2+1-0.3,-1.65) {$M^{(\U)},  M^{(\e)}$};
\draw [ very thick, ->] (-2.4, -1.75+1)--(-1.5, -1.75+1);
\node[draw =none] at (-2,-1.65+1.1) {$ M^{(\e)}$};
\node[draw =none] at (-1.1,-1.75+1) {Enc};
\node[draw =none] at (-1.1+1,-1.75-0.2) {Enc};
\node[draw =none] at (-0.5,-0.6) {$\vect X_1$};
\node[draw =none] at (-0.5+1,-0.6-0.75) {$\vect X_2$};
\node[draw =none] at (-0.5+1,-0.6-0.75-0.75) {$\vect X_3$};
\draw (3.3,-1.75+0.75) rectangle (4.05,-1.25+0.75);
\node[draw =none] at (-0.5+2.25,-0.6) {$\vect Y_1$};
\node[draw =none] at (3,-0.6-0.75) {$\vect Y_2$};
\node[draw =none] at (4.5,-0.6-0.75-0.75) {$\vect Y_3$};
\draw (1.5+0.5+1.3+1.5,-1.75+0.75-0.75) rectangle (5.55,-1.25+0.75-0.75);
\draw [ very thick, ->] (1.5, -0.75)--(3.3, -0.75);
\draw [ very thick, ->] (2.75, -1.5)--(4.8, -1.5);
\draw [ very thick] (4.25, -2.25)--(5.2, -2.25);
\draw [ very thick, ->] (5.2, -2.25)--(5.2,-1.75);
\draw [ very thick, ->] (3.6, -1.5)--(3.6,-1);
\draw [ very thick, ->] (4.05, -0.75)--(4.5, -0.75);
\draw [ very thick, ->] (5.55, -1.5)--(6.2, -1.5);
\node[draw =none] at (4.65+0.1+0.2,-0.75) {$\hat M^{(\e)}$};
\node[draw =none] at (6.4+0.2,-1.5) {$\hat M^{(\U)}$};
%\draw [ very thick, ->] (5.2, -0.3)--(5.2,-1.25);
%\draw [ very thick] (5.2, -0.3)--(2.5,-0.3);
%\draw [ very thick] (2.5, -0.75)--(2.5,-0.3);
\node[draw =none] at (0.75,-0.75) {$P_{\vect Y_1| \vect X_1}$};
\node[draw =none] at (2.1,-1.5) {$P_{\vect Y_2| \vect X_2}$};
\node[draw =none] at (3.5,-2.25) {$P_{\vect Y_3| \vect X_3}$};
\node[draw =none] at (3.7,-0.75) {Dec};
\node[draw =none] at (5.2,-1.5) {Dec};
\draw[very thick, ->] (-1.5,-4.5)--(5.2,-4.5);
\draw[very thick, ->] (-1.5,-4.5)--(-1.5,-3.25);
\draw[very thick, ->, green] (-0.5,-4)--(-0.5,-4.5);
\draw[very thick, ->, green] (3,-4.5)--(3,-5);
\draw [ very thick, ->, red] (1.5,-4)--(1.5,-4.5);
\draw [ very thick, ->, red] (4.5,-4.5)--(4.5,-5);

\node[draw =none] at (-0.5+0.1,-3.8) {$M^{(\e)}$};
\node[draw =none] at (-0.5,-4.8) {$t = a_{\e}$};
\node[draw =none] at (1.5,-3.8) {$M^{(\U)}$};
\node[draw =none] at (1.5,-4.8) {$t = a_{\U}$};
\node[draw =none] at (3.1,-3.8-1.3-0.1) {$\hat M^{(\e)}$};
\node[draw =none] at (3,-4.3) {$t = d_{\e}$};
\node[draw =none] at (4.6,-3.8-1.3-0.1) {$\hat M^{(\U)}$};
\node[draw =none] at (4.5,-4.3) {$t = d_{\U}$};
\node[draw =none] at (5.4,-4.5) {$t$};
\end{tikzpicture}
  \caption{System model for transmission of URLLC and eMBB messgaes in the finite blocklength regime and under heterogeneous decoding deadline.}
   \label{fig2} 
  \end{figure}
This section is based on the results in \cite{WCNC2021long} and \cite{ISIT2022long}. We again consider a P2P scenario, but where  communication is over a non-fading Gaussian channel  
\begin{equation*}
Y_{t} = X_t +Z_t,\qquad t=1,2,\ldots, 
\end{equation*}
for $\{Z_t\}$ an i.i.d. standard Gaussian noise sequence. %The channel input sequence is subject to an average block-power constraint $\P$. 

The Tx has a single URLLC message and a single eMBB message to send to the Rx, where both messages are assumed to have strict creation times and fixed decoding deadlines. Specifically, transmission of eMMB Message $M^{(\e)}$ commences at time  $t = a_\e$ and decoding has to performed at time $t=d_\e$, while URLLC message can be transmitted starting at time $t=a_{\U}$ and has to be decoded at time $t=d_{\U}$. We thus parametrize the message sets as $\mathcal{M}_{\U}=\big[2^{(d_{\U}-a_{\U})R_{\U}}\big]$ and  $\mathcal{M}_{\e}=\big[2^{(d_{\e}-a_{\e})R_{\e}}\big]$.
We also denote the transmission window of the eMBB message by $\mathcal{W}_{\e}$ and the  transmission window of the URLLC messages by $\mathcal{W}_{\U}$:
\begin{align}
\mathcal{W}_{\e} \overset{\triangle}{=} \{a_{\e},\ldots,d_{\e}\},~~~\mathcal{W}_{\U} \overset{\triangle}{=} \{a_{\U},\ldots,d_{\U}\}.
\end{align}
Since the URLLC delay  $d_{\U}-a_{\U}$ is assumed shorter than the eMBB delay $d_{\e}-a_{\e}$, the following three situations can arrive: 
\begin{itemize}
\item Case $1$:    URLLC  and eMBB transmissions do not overlap. i.e.,  $\mathcal{W}_{\U} \cap \mathcal{W}_{\e}=\emptyset$.
%\begin{equation*}
%a_{\e} < d_{\e} < a_{\U} < d_{\U} \qquad \textnormal{or}\qquad  a_{\U} < d_{\U} < a_{\e} < d_{\e} .
%\end{equation*} 
\item Case $2$: The eMBB transmission interval includes the URLLC  transmission interval, i.e., $\mathcal{W}_{\U}  \subset \mathcal{W}_{\e}$.
%\begin{equation*}
%a_{\e}  < a_{\U} < d_{\U}  < d_{\e} .
%\end{equation*} 
 %p. i.e., either   the transmission of the eMBB message overlaps  the transmission of the URLLC message, i.e.,  
%More specifically, the URLLC message arrives while the Tx is sending the eMBB message and needs to be decoded after 
\item Case $3$:  URLLC and eMBB transmissions overlap, but URLLC transmission is not included in eMBB transmission, i.e.,  $\mathcal{W}_{\e} \cap \mathcal{W}_{\U} \neq \emptyset $ and $\mathcal{W}_{\U} \not \subset \mathcal{W}_{\e}$.
%\begin{equation*}
%a_{\e}  < a_{\U} < d_{\e}< d_{\U} \qquad \textnormal{or} \qquad a_{\U} < a_{\e} <d_{\U}  < d_{\e} .
%\end{equation*}
 % the transmission of the URLLC message commences after the arrival of the  eMBB message and terminates before the decoding time of  the eMBB message, i.e., $\mathcal{W}_{\U} \subset\mathcal{W}_{\e}$. %, then % transmission of the eMBB message and has the b
\end{itemize}
In  Case $1$ where the two messages are transmitted during independent time intervals, URLLC and eMBB transmissions can be analyzed independently based on the  
%upper and lower bounds on the error probability of each message c new characterizations of fundamental trade-offs between the size of the message set, the probability of error, and the length of the code can be obtained
% via
achievability and converse bounds  in \cite{Yuri2012}.  Cases 2 and 3 can be treated similarly. Here, we focus on a subcase of  Case 3  where encoding starts with the eMBB message at time $t=a_{\e}$ and terminates at time $t=d_{\U}$ with the decoding of the URLLC message,  see Figure~\ref{fig2}. The Tx thus produces %Channel inputs are thus formed as
%
%In \cite{WCNC2021long} and \cite{ISIT2022long}, however the focus is on Case $2$ where the transmissions of two messages overlap. 
%It is easy to cover Case $3$ with the analysis provided in \cite{WCNC2021long} and \cite{ISIT2022long}. In Case $2$, the encoder outputs
% symbols 
channel inputs at times $t \in \{a_{\e},\ldots,d_{\U}\}$ as follows: 
\begin{align}
X_t = \begin{cases} {f}_t ( M^{(e)}), & \quad t \in \{a_{\e}, \ldots,a_{\U}-1\}\\
\psi_t(M^{(e)}, M^{(U)}), & \quad t \in \{a_{\U}, \ldots, d_{\e}\} \\
\phi_t(M^{(U)}), & \quad t \in \{d_{\e} + 1, \ldots, d_{\U}\},
\end{cases} 
\end{align}
where $\{f_t\},\{\psi_t\},\{\phi_t\}$ are appropriate encoding functions. % the encoding functions corresponding to the channel uses where only message $ M^{(\e)}$ is arrived but not $M^{(\U)}$, where both $ M^{(\e)},M^{(\U)}$ are present, and after $ M^{(\e)}$ has been decoded, respectively. 
Note that the Tx does not know the URLLC message before time $t = a_{\U}$ and therefore channel inputs prior to time $t=a_{\U}$ cannot depend on $M^{(\U)}$. It can  also be assumed that channel inputs after the eMBB decoding time $d_{\e}$ de not depend on the eMBB message $M^{(\e)}$. One can therefore think of the transmission taking place over three parallel channels, with respective blocklengths 
\begin{equation}
n_1 \triangleq a_{\U} - a_{\e}, \quad n_{2} \triangleq d_{\e} - a_{\U}+1, \quad \text{and} \quad n_3 \triangleq d_{\U} - d_{\e},
\end{equation}
 where the first channel consists of channel uses $\{a_{\e},\ldots, a_{\U}-1\}$ and  incorporates only eMBB transmission; the second channel consists of  channel uses  $\{a_{\U},\ldots, d_{\e}\}$ and incorporates \emph{joint transmission} of URLLC and eMBB messages; and  the third channel consists of  channel uses $\{d_{\e}+1,\ldots, d_{\U}\}$ and incorporates only URLLC transmission. 
%The goal is to establish bounds on the decoding error probabilities of the eMBB message $M^{(e)}$  and the URLLC message $M^{(U)}$ in the finite blocklength regime.% under a Gaussian interference assumption defined precisely in Section~\ref{sec:main}. 
We denote the inputs and outputs of the three parallel  channels by 
\begin{subequations}
	\begin{IEEEeqnarray}{rClrCl}
		\vect X_1 &\triangleq& \{X_{a_{\e}}, \ldots, X_{a_{\U}-1}\},  \qquad& \vect Y_1 &\triangleq& \{Y_{a_{\e}}, \ldots, Y_{a_{\U}-1}\}, \\
		\vect X_2&\triangleq& \{X_{a_{\U}}, \ldots, X_{d_{\e}}\},\qquad &\vect Y_2&\triangleq& \{Y_{a_{\U}}, \ldots, Y_{d_{\e}}\},\ \\
		\vect X_3&\triangleq& \{X_{d_{\e}+1}, \ldots, X_{d_{\U}}\}, \qquad &\vect Y_3&\triangleq& \{Y_{d_{\e}+1}, \ldots, Y_{d_{\U}}\}. 
	\end{IEEEeqnarray}
\end{subequations}
% For each $i \in \{1, 2,3\}$, define $P_i$ as the average power associated over the $i$-th channel of $n_i$ channel uses.
%{For the $i$-th channel with $i \in \{1, 2,3\}$, the encoding functions satisfy the average block power constraint
% \begin{IEEEeqnarray}{rCl}
%\frac{1}{n_i}  ||\vect X_i||^2 \le  P_i \IEEEeqnarraynumspace
%\end{IEEEeqnarray}
%almost surely.}
% Assume that codewords have constant power in each channel, that is $\vect X \in \mathcal K$ with set $\mathcal K$ defined as
% \begin{IEEEeqnarray}{rCl}
%\mathcal K = \left \{ \vect X = [ \vect X_1, \vect X_2, \vect X_3] \Big | ||\vect X_i||^2 = n_i P_i \right \}. \IEEEeqnarraynumspace
%\end{IEEEeqnarray}
For the $i$-th channel with $i \in \{1, 2,3\}$, the encoding functions satisfy the average block power constraint
 \begin{IEEEeqnarray}{rCl}
\frac{1}{n_i}  ||\vect X_i||^2 \le  \P_i \IEEEeqnarraynumspace
\end{IEEEeqnarray}
almost surely.
The resulting system model is illustrated in Figure~\ref{fig2}, where notice that  the three   channels  $P_{\vect Y_1| \vect X_1}$, $P_{\vect Y_2| \vect X_2}$ and $P_{\vect Y_3| \vect X_3}$ are additive, memoryless, stationary, and Gaussian of variances $1$.

%The scheme in \cite{WCNC2021long} splits the available average block-power $\P$ into powers $\P_1, \P_2, \P_3$ summing to $\P$, where each  $\P_i$ denotes the average block-power  dedicated to channel  $i$, for  $i\in\{1,2,3\}$. 
The scheme further proposes  to combine the eMBB and URLLC transmission over the second channel $P_{\vect Y_2| \vect X_2}$ by means of the simple superposition coding approach described at the end of Subsection~\ref{sec:superpos}, for which the block-2 power $\P_2$ is split into power $\beta \P_2$ for the eMBB  transmission and power $(1-\beta_2)\P_2$ for the URLLC transmission, for some $\beta \in [0,1]$. 
In particular, the channel inputs $\vect{X}_2$ are formed as 
\begin{equation}
\vect{X}_2=\vect{X}_{2,\e} +\vect{X}_{2,\U},
\end{equation}
where $\vect{X}_{2,\e}$ is a codeword encoding $M^{(\e)}$  of average power $\|\vect{X}_{2,\e} \|^2 = n_2 \beta \P_2$, and $\vect{X}_{2,\U}$ is a codeword encoding $M^{(\U)}$  of average power $\|\vect{X}_{2,\U} \|^2 = n_2 (1-\beta) \P_2$.

The Rx first decodes the eMBB message based on the outputs of the first and second channels where it  treats the transmission of the URLLC message over the second channel as interference. Subsequently, it decodes the URLLC message based on the outputs of the second and third channel, conditioning on the already decoded eMBB message. 

The error probabilities of the described scheme can be analyzed and compared to fundamental lower bounds on the error probabilities, obtained via \emph{meta-converse} arguments \cite{Yuri2012} with an extension to  parallel channels \cite{Erseghe2016}.  As shown  in \cite{WCNC2021long},   the converse and achievability bounds match in specific cases.

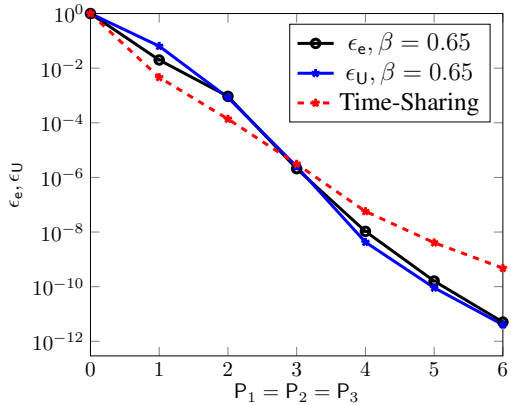
\begin{figure}[t!]
\centering
\begin{tikzpicture}[scale=0.8]
\begin{axis}[
    xlabel={\small {$ \P_1 =\P_2 = \P_3$ }},
    ylabel={\small {$\epsilon_{\e}, \epsilon_{\U}$ }},
     xlabel style={yshift=.5em},
     ylabel style={yshift=0em},
    xmin=0, xmax=6,
    ymin=0, ymax=1,
    xtick={0,1,2,3,4,5,6},
    ytick={0,1e-16, 1e-14,1e-12, 1e-10,1e-8,1e-6,1e-4, 1e-2,1},
    yticklabel style = {font=\small,xshift=0.25ex},
    xticklabel style = {font=\small,yshift=0.25ex},
    legend pos=north east,
    ymode=log,
]

 \addplot[ color=black,   mark=halfcircle, line width = 0.5mm] coordinates {  (0,1)(1,0.019907012386709)(2,9.29815901371534e-04)(3,2.08913272237155e-06)(4,1.05647601778003e-8)(5,1.606270672027676e-10)(6,0.05031905536633e-10)};

\addplot[ color=blue,   mark=star, line width = 0.5mm] coordinates {(0,1)(1,0.0640331830595192)(2,8.40165247509850e-04)(3,2.51158031039367e-06)(4,4.26914059659111e-9)(5,0.910516437917067e-10)(6,0.040216941716030e-010)};

%\addplot[ color=green,   mark=star, line width = 0.5mm, dashed] coordinates {(0,1)(1,0.004594334957071)(2,0.000013738239466)(3,0.000000002097657)(4,0.000000002714645)(5,0.000000060804159)(6, 0.000475296645264e-03)};
\addplot[ color=red,   mark=star, line width = 0.5mm, dashed] coordinates {(0,1)(1,0.004594334957071)(2,0.00013738239466)(3,0.000003097657)(4,0.00000005714645)(5,0.0000000040804159)(6, 0.00475296645264e-07)};

\legend{{$\epsilon_{\e}, \beta = 0.65$}, {$\epsilon_{\U}, \beta = 0.65$}, {Time-Sharing} }  
\end{axis}

\vspace{-0.4cm}
\end{tikzpicture}

\caption{The figure illustrates the average error probabilities of the eMBB and URLLC messages denoted by $\epsilon_{\e}$ and $\epsilon_{\U}$ in function of the block transmit powers $\P_1 = \P_2 = \P_3$ and for blocklengths $n_1 = 20, n_2 = 20, n_3 = 20$, URLLC rate $R_{\U} = 1/4$, and power split $\beta = 0.65$.}
\label{fign2}
\vspace{-0.5cm}
\end{figure} 
%The numerical analysis of the converse and achievability bounds obtained in \cite{}  is shown  in Fig.~\ref{fign1} where the bounds on $\epsilon_1$ and $\epsilon_2$ are evaluated as a function of the transmit power for different values of the parameter $\beta$. %We assume equal transmit power over the all three channels with $n_1 = n_2 = n_3 = 10$ and the Gaussian interference approximation holds. Note that utilizing Corollary~\ref{col1}, the upper and lower bounds are in agreement.
 
%In the second channel recall that $\beta P_2$ is the power assigned to  transmit $m_1$ and $(1-\beta)P_2$ is the power assigned to transmit $m_2$. Observe that as the power sharing parameter $\beta$ increases, the error probability $\epsilon_1$ decreases and $\epsilon_2$ increases, as expected. When $\beta = 0.5$, i.e., when the transmit power $P_2$ is assigned equally to the transmission of each message, $\epsilon_2$ is lower than $\epsilon_1$. This is due to the fact that when decoding the first message, the transmission of the second message is considered as interference. On the other hand, when decoding the second message, the first message has already been decoded. 
 
 Fig.~\ref{fign2}, from \cite{WCNC2021long},   compares the performance  of the described superposition coding scheme with a standard scheduling scheme  that allocates the first half of the channel uses to eMBB transmission and the second half to URLLC transmission.  (Under this scheduling approach $\epsilon_{\e}=\epsilon_{\U}$.)  One observes that for the chosen set of parameters, $n_1= n_2 = n_3 = 20$ $R_{\U}=1/4$,   and $\beta=0.65$, the superposition coding approach results in almost identical URLLC and eMBB error probabilities $\epsilon_{\e}$ and $\epsilon_{\U}$.  Moreover, at medium and high  powers $P_2$, the superposition coding approach significantly outperforms  the scheduling approach. 

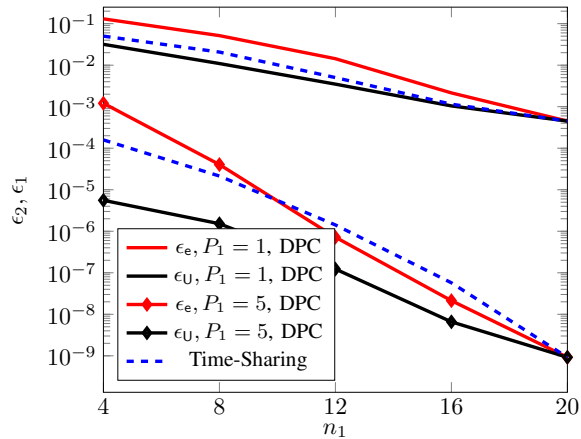
\begin{figure}[t!]
\centering
\begin{tikzpicture}[scale=0.9]
\begin{axis}[
    xlabel={\small {$n_1$ }},
    ylabel={\small {$\epsilon_2, \epsilon_1$ }},
     xlabel style={yshift=.5em},
     ylabel style={yshift=0em},
    xmin=4, xmax=20,
    ymin=0, ymax=0.25,
    xtick={4,8,12,16,20},
    ytick={0,1e-9,1e-8,1e-7,1e-6,1e-5,1e-4,1e-3,1e-2,1e-1,1},
    yticklabel style = {font=\small,xshift=0.25ex},
    xticklabel style = {font=\small,yshift=0.25ex},
    legend pos=south west,
    ymode=log,
]

 \addplot[ color=red, line width = 0.5mm] coordinates {  (4,0.13060527692151)(8,0.051508342663946)(12,0.014328665066819)(16,0.002149441086310)(20,0.00045)};

\addplot[ color=black,  line width = 0.5mm] coordinates {  (4,0.031797749837211)(8,0.010954079530531)(12,0.003496016228159)(16,0.001045204309634)(20,0.00045)};

 \addplot[ color=red,   mark=diamond, line width = 0.5mm] coordinates {  (4,0.00120983577087)(8,0.000040421691731)(12,0.000007196125933e-1)(16,0.00000021315578850624e-1)(20,0.922e-9)};

 \addplot[ color=black,   mark=diamond, line width = 0.5mm] coordinates {  (4,0.0000556552184607e-1)(8,0.000015191722545e-1)(12,0.0000012297995561e-1)(16,0.00000006586198814945e-1)(20,0.922e-9)};

%\addplot[ color=blue,  dashed , line width = 0.5mm] coordinates {  (4,0.0108516728095829e-2)(8,0.151372279230886e-4)(12,0.0111893571374336e-4)(16,0.00000005739265)(20,0.922e-9)};

 \addplot[ color=blue,  dashed , line width = 0.5mm] coordinates {  (4,0.500898338783796e-1)(8,0.206920410722971e-1)(12,0.0502842595317840e-1)(16,0.001149441086310)(20,0.00045)};

\addplot[ color=blue,  dashed , line width = 0.5mm] coordinates {  (4,0.0158516728095829e-2)(8,0.21372279230886e-4)(12,0.0141893571374336e-4)(16,0.00000005739265)(20,0.922e-9)};

%\legend{ {\footnotesize $\epsilon_2, P_1= 1$}, {\footnotesize$\epsilon_1, P_1  = 1$}, {\footnotesize TS,$P_1 = 1$}, {\footnotesize$\epsilon_2, P_1= 5$}, {\footnotesize$\epsilon_1, P_1  = 5$}, {\footnotesize TS,$P_1 = 5$}}  

\legend{ {\footnotesize $\epsilon_{\e},P_1= 1$, DPC}, {\footnotesize $\epsilon_{\U}, P_1  = 1$, DPC}, {\footnotesize$\epsilon_{\e}, P_1= 5$, DPC}, {\footnotesize$\epsilon_{\U}, P_1  = 5$, DPC}, {\footnotesize Time-Sharing}}  
\end{axis}

\vspace{-0.8cm}
\end{tikzpicture}

\caption{DPC based upper bounds on  $\epsilon_{\U}$ and $\epsilon_{\e}$  in function of the blocklength $n_1$ and for average block-powers $\P_1 = \P_2 = \P_3 $.} %, $\beta_{\e} = 0.4$ and $\beta_{\U} = 0.4$. }%, $\kappa = 1$, $\epsilon_T = 0.000001$ }
\label{fig7}
\vspace{-0.5cm}
\end{figure}

In \cite{ISIT2022long}, the authors extend above coding scheme by using the finite-blocklength  dirty paper coding (DPC)  scheme in \cite{Scarlett2015} to  precancel the interference of the eMBB message on the URLLC transmission. Notice that in  finite-blocklength DPC, the joint-typicality check is replaced by a norm condition on the input signal, and the Tx has to sacrifice few channel uses to approximately describes the norm of the interference sequence to the Rx. 
The  error probabilities of this DPC based scheme are analyzed in \cite{ISIT2022long} based on the DPC analysis technique  in \cite{Scarlett2015} and the parallel channel extension analysis in \cite{Erseghe2016}. %The DPC coding proposed in \cite{Scarlett2015} uses the 
Figure~\ref{fig7}, from  \cite{ISIT2022long},    compares the performances of the proposed DPC based scheme with standard scheduling, and shows that for large transmit powers $\P_1=\P_2=\P_3$, the DPC based scheme outperforms scheduling over a wide range of blocklengths $n_1$. 

%  the probability density function  
%\begin{IEEEeqnarray}{rCl}\label{eq:fu}
%	f_{\vect U} ^{(\pt)}(\vect u) = \frac{\delta \left (||\vect u||^2 - n_{2,2} (\beta_{\e} P_2 + \alpha^2\pt)\right )}{S_{n_{2,2}} (\sqrt{n_{2,2} (\beta_{\e} P_2 + \alpha^2 \pt) })},
%\end{IEEEeqnarray}
%where $\delta(\cdot)$ is the Dirac delta function, and $S_{n}(r) $ is the surface area of a sphere of radius $r$ in $n$-dimensional space.
 
\subsection{Summary}\label{sec:P2P_summary}
This section  considered a P2P channel with a single Tx that sends both an URLLC and an eMBB message, where the two types of messages have different decoding delays. In Subsection~\ref{sec:superpos}, transmission is over fading channels and URLLC messages have to be transmitted within a single coherence block, whereas eMBB messages can be sent over multiple blocks and thus profit from channel diversity. To compensate for the missing channel state-information at the Tx, an infinite-layer broadcast approach is employed. A closed-form solution for 
the  sum-rate achieved by this broadcast approach was presented, and based on numerical simulations it was observed its maximum sum-rate  decreases with increasing URLLC rates. Furthermore, a simplified single-layer power allocation was shown to perform close to the optimal power allocation in the broadcast approach at low URLLC rates. 

Subsection~\ref{sec:consecutive} studies a related simplified superposition coding or dirty-paper coding schemes but over static Gaussian channels and with fixed creation and decoding times. For this setup, upper and lower bounds on the set of achievable  error probability pairs that can simultaneously be achieved for URLLC and eMBB messages  was derived in \cite{WCNC2021long}. The obtained results show a performance improvement under these schemes compared to the standard scheduling scheme.

\section{Broadcast Channels with Mixed-Delay Traffic}\label{sec:BC}
\subsection{Introduction}
This section focuses on multi-receiver broadcast channels (BC).  The results of this section are based on \cite{Lin2021} where similarly to Section~\ref{sec:superpos}, URLLC messages have to be decoded within a single coherence block but eMBB messages can be transmitted over multiple blocks. In contrast to Section~\ref{sec:superpos}, {the fading powers $\{S_{k,b}\}$ of the various blocks  are known to the various Rxs and the Tx in advance}. 
% Subsection~\ref{sec:superpos} reviews the superposition coding approach  over fading channels in  \cite{shlomo2012ISIT, Cohen2012}. This approach manages to send URLLC messages over single coherence blocks of the fading channel without suffering from a degradation due to the lack of state knowledge at the Tx,  and to simultaneously send eMBB messages over multiple coherence blocks, exploiting the ergodic behaviour of the channel. The  broadcast approach is a general framework to combine different types of transmissions and was discussed in detail in a recent survey paper \cite{Tajer2021}.  
%Subsection~\ref{sec:consecutive} summarizes the results in \cite{WCNC2021long}, \cite{ISIT2022long}, which analyze a similar approach  for Gaussian channels and in the finite-blocklength regime. The analysis is based on the concept of “parallel channels” introduced in \cite{Erseghe2016} which distinguishes the transmission intervals  with and without URLLC. 
%In this section we consider  multi-receiver broadcast channels (BC). Two lines of work have been proposed ...

\subsection{Broadcast Approach over Fading Channels} \label{sec:BCfading}
% where, similarly to Section~\ref{sec:superpos}, URLLC messages have to be decoded within a single coherence block but eMBB messages can be transmitted over multiple blocks. In contrast to Section~\ref{sec:superpos}, we assume that the fading powers $\{S_{k,b}\}$ in the various blocks and to the various receivers are known in advance by the recei and all receivers. Moreover, in this setup, each receiver might demand a URLLC message, an eMBB message or both. We thus define the two sets  $\mathcal{K}^{(\U)}$ and  $\mathcal{K}^{(\e)}$ indicating the sets of users requesting URLLC and eMBB messages, respectively. Notice that the two sets can overlap. 
%\subsection{}
 { In the setup proposed in \cite{Lin2021} each Rx might demand a URLLC message, an eMBB message or both. We thus define the two sets  $\mathcal{K}^{(\U)}$ and  $\mathcal{K}^{(\e)}$ indicating the sets of users requesting URLLC and eMBB messages, respectively. Notice that the two sets can overlap}. The mixed-delay constraint is captured by imposing a fixed rate on all transmitted URLLC messages, whereas eMBB messages can be either of larger or smaller rates. This rate-adaption on eMBB messages depending on the encountered fading powers allows to increase the system's sum-rate. The corresponding optimization problem can be expressed as 
\begin{equation}\label{eq:optim} 
\max  \sum_{k \in \mathcal{K}^{(\U)} }  R^{(\U)}+  \sum_{k \in \mathcal{K}^{(\e)} }  R_k^{(\e)} ,
\end{equation}
where the maximization is only over rate-tuples such that URLLC rates $\{R_{k,b}^{(\U)}=R^{(U)}\}_{k\in \mathcal{K}^{(\U)}}$ are achievable on each block $b\in\{1,\ldots, B\}$ and eMBB rates $\{R_k^{(\e)}\}_{k\in\mathcal{K}^{(\e)}}$ are simultaneously achievable over the entire transmission, all using  dirty-paper coding with an optimal precoding order and under an average block power constraint $\P$.

Optimization problem \eqref{eq:optim} is cumbersome to solve, and instead \cite{Lin2021} proposes the following suboptimal algorithm. Fix the dirty-paper precoding order to first precode the eMBB messages followed by the URLLC messages. This implies that eMBB transmissions act as noise on the URLLC communication but not vice versa. Then choose a target URLLC rate $R^{(U)}$ and find the minimum required average block-power $\beta\P$, for $\beta \in[0,1]$, that ensures achievability of the per-block and per-user URLLC rate $R^{(\U)}$. Identify finally the maximum sum-rate $ \sum_{k \in \mathcal{K}^{(\e)} } R_k^{(\e)}$ achievable on the eMBB transmission with average power $(1-\beta)\P$. 

Though optimal, dirty-paper coding is difficult to implement in practical systems and is often replaced by the simpler  zero-force beamforming. In the context of our multi-user and mixed-delay communication scenario, under zero-force beamforming, it remains to determine the assignment of  beams to users and the two  communication types. The work in \cite{Lin2021} proposes a sophisticated beam assignement algorithm, which assigns stronger sub-channels (beams) to URLLC messages, and weaker channels to eMBB messages. The idea being that eMBB communication can profit from channel diversity over multiple coherence blocks. 

Numerical simulations in \cite{Lin2021} compare the sum-rate in \eqref{eq:optim} achieved with dirty-paper coding and with a precoding order and power allocation established according to the suboptimal algorithm described above, with the sum-rate achieved with a  beamforming alternative. For both schemes the maximum sum-rate increases with small values of $R^{(\U)}$ and reaches a peak when the URLLC rate contributes approximately a third of the sum-rate. Beyond, the sum-rate decays rapidly because the delay constraint on the URLLC message becomes too stringent and limits the overall performance.

\subsection{Summary} 
This section  extended the superposition coding approach to fading BCs with multiple Rxs where certain Rxs demand URLLC messages and other eMBB messages. Assuming perfect channel state information, \cite{Lin2021}  proposes  precoding orders or beam assignments for URLLC and eMBB messages, that take into account that URLLC messages have to achieve their desired rates in a single coherence block and therefore cannot exploit the channel diversity offered over multiple blocks. The results in \cite{Lin2021} show that for small requested URLLC rates, the sum-rate of the system is not limited by the stringent delay constraint of URLLC messages. For larger URLLC rates this is however the case and URLLC delay constraints limit the overall performance.

%%%%%%%%%%%%%%%%%%%%%%%%%%%%%%%

\section{Cooperative Interference Networks} \label{sec:Network}
\subsection{Introduction} 
In this section we consider interference networks where Txs and/or Rxs can cooperate over dedicated cooperation links. This models for example cellular networks where BSs can cooperate over high-rate  fiber-optic links, and neighbouring mobiles can cooperate using bluetooth or millimeter wave communication, which take place on different frequency bands than the standard radio communication between mobiles and BSs,  and  cause no interference.

\begin{figure}[h!]
 \begin{tikzpicture}[scale=1.35, >=stealth]
 \centering
\draw[thick, ->] (0,3.75)--(10.6,3.75);
%\draw[thick] (-0.75,3.65)--(-0.75,3.85);
{\node [draw=none] at (10.8, 3.75) {$ t$};}
%%%channel 
\draw[line width= 1mm, black] (4,3.75)--(5.5,3.75);
\draw[thick] (4,3.65)--(4,3.85);
\draw[thick] (5.5,3.65)--(5.5,3.85);
{\node [draw=none] at (4, 4) {$ t_0$};}
{\node [draw=none] at (5.5, 4) {$ t_0+n$};}
{\node [draw=none] at (4.75, 3.5) {channel};}
%{\node [draw=none] at (4, 3.8) {interference network};}

% Cooperation blocks
\draw[line width= 2mm, gray] (1,3.75)--(4,3.75);
\draw[line width= 2mm, gray] (5.5,3.75)--(10,3.75);

{\node [draw=none] at (2.5, 3.4) {Tx-cooperation rounds};}
{\node [draw=none] at (2.5, 3.1) {$1,\ldots, \Dt$};}

{\node [draw=none] at (7.75, 3.4) {Rx-cooperation rounds};}
{\node [draw=none] at (7.75, 3.1) {$1,\ldots, \Dr$};}

{\node [draw=none] at (2.5, 4) {$ t_0-n$};}
% Block -1
{\node [draw=none] at (2.5, 4) {$ t_0-n$};}
\draw[thick] (2.5,3.65)--(2.5,3.85);

% Block -2
{\node [draw=none] at (1, 4) {$ t_0-2n$};}
\draw[thick] (1,3.65)--(1,3.85);

%% Block -3
%{\node [draw=none] at (-0.5, 4) {$ t_0-3n$};}
%\draw[thick] (-0.5,3.65)--(-0.5,3.85);

% Block +2
{\node [draw=none] at (7, 4) {$ t_0+2n$};}
\draw[thick] (7,3.65)--(7,3.85);

% Block  +3
{\node [draw=none] at (8.5, 4) {$ t_0+3n$};}
\draw[thick] (8.5,3.65)--(8.5,3.85);

% Block +4
{\node [draw=none] at (10, 4) {$ t_0+4n$};}
\draw[thick] (10,3.65)--(10,3.85);

% arrivals of URLLC 
\draw[line width= 0.5mm, <-, red] (3.9,3.75)--(3.9,4.2);
\draw[line width= 0.5mm, <-, red] (3.7,3.75)--(3.7,4.2);
\draw[line width= 0.5mm, <-, red] (3.1,3.75)--(3.1,4.2);

%decoding of URLLC
\draw[line width= 1.4mm, ->, red] (5.5,3.75)--(5.5,3.3);
{\node [draw=none] at (5.5, 3.1) {URLLC};}
{\node [draw=none] at (5.5, 2.8) {decoding};}

% arrivals of eMBB
\draw[line width= 0.5mm, <-, blue] (0.2,3.75)--(0.2,4.2);
\draw[line width= 0.5mm, <-, blue] (0.5,3.75)--(0.5,4.2);
\draw[line width= 0.5mm, <-, blue] (0.9,3.75)--(0.9,4.2);

%decoding of eMBB

\draw[line width= 1.4mm, ->, blue] (10,3.75)--(10,3.3);
{\node [draw=none] at (10, 3.1) {eMBB};}
{\node [draw=none] at (10, 2.8) {decoding};}

\end{tikzpicture}
\vspace{.1cm}
\caption{Timeline of cooperation and transmission over the interference network  for URLLC and eMBB messages associated to the block from time $t_0$ to time $t_0+n$.}
\label{fig:timeline}
\end{figure}
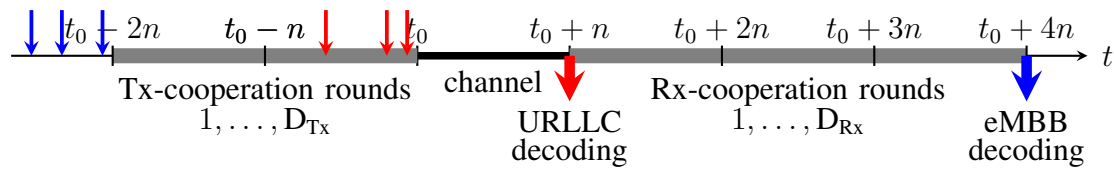

Cooperation links between Txs are beneficial for eMBB transmissions, because they allow  Txs to exchange parts of their messages or their signals so as to enable cooperative signaling over the channel. Cooperation links between Rxs can be used  to exchange information about  receive signals or   decoded messages,  allowing the Rxs to better mitigate  interference. URLLC  transmissions however have to start immediately after the  creation of the messages  and the additional delays caused by exchanging (parts of) URLLC messages between Txs cannot be tolerated. In the same sense,  URLLC messages have to be decoded before Rxs can learn information about other Rxs' decoded messages or receive signals. 
Figure~\ref{fig:timeline} illustrates a typical timeline  in our model. The actual communication time over the interference network is from time $t_0$ to time $t_0+n$ and corresponds to the blocklength of communication. Here $t_0$ denotes an arbitrary starting time of a block and $n$ refers to the block length. It  is dedicated to the  transmission of URLLC messages generated just prior to $t_0$ and of eMBB messages generated prior to $t_0- \Dt \cdot n$, so as to allow the eMBB messages to profit from $\Dt$ rounds of Tx-cooperation. (For simplicity it is assumed that $n$ represents also the length of a cooperation round. The results also extend to scenarios where this is not the case.) Rxs decode the URLLC messages transmitted in this block $[t_0,t_0+n]$ as soon as the block is terminated, each Rx simply based on its receive signal. Decoding of eMBB messages can be delayed to time $t_0+(\Dr+1) \cdot n$, until the termination of $\Dr$ Rx-cooperation rounds. %In a similar way, during any other block  $[t_0+\kappa n, t_0+(\kappa+1)n]$, for $\kappa$ a non-zero integer, communication over the interference network handles the eMBB messages created just before time $t_0+(\Dt-\kappa)n$ (so that they can profit from $\Dt$ Tx-cooperation rounds) and the recently created URLCC messages just before time $t_0+\kappa n$. The URLLC messages are directly decoded at the end of the block, at time $t_0+(\kappa+1)n$, and decoding of  eMBB messages can be delayed to after $\Dt$ Rx-cooperation rounds, i.e., to time $t_0+(\kappa +\Dr+1)n$.

Consider a scenario where each Tx  sends an URLLC message and an eMBB message to its corresponding Rx. The focus is on the set of \emph{degrees of freedom (DoF)} pairs that are simultaneously achievable for URLLC and eMBB messages, and in particular on how the sum-DoF decreases with increasing URLLC DoFs. This decrease describes the degradation of the overall system performance  caused by the stringent delay constraints on URLLC messages, as a function of the URLLC rates.  Somehow surprisingly, it can be shown that such a degradation does not exist in a variety of networks even with moderate or large URLLC DoFs. 

In the following Subsection~\ref{setup} we describe the problem setup. In Subsection~\ref{sec:coding}, we  present the integrated scheduling and  coding scheme for  URLLC and eMBB messages in \cite{HomaTCOM2021}, and in Subsection~\ref{sec:results} we show that this scheme achieves  maximum sum-rate even for moderate or large  URLLC rates  on a variety of network topologies, thus limiting the degradation of the overall system performance. % delay constraints of URLLC communication on the overall performance of the system. 
In Subsection~\ref{sec:random} we discuss  a random-arrival model for URLLC and eMBB messages, where URLLC and eMBB messages are assigned to users according to some random arrival process. Again based on the coding scheme in \cite{HomaTCOM2021},  it can be shown that even under random arrival messages, the overall system performance is hardly degraded by the strict URLLC delay constraints \cite{HomaITW2020, HomaITW2021}.

\renewcommand{\K}{\mathsf{K}}

\subsection{Problem Description} \label{setup}

Throughout this section, we consider a cellular network, but  assume that users of the same cell are scheduled in different frequency bands. Interference thus occurs only from  the mobile users in neighbouring cells that are scheduled on the same frequency band. The system therefore decomposes into subsystems with only a single mobile in each cell. 

 Consider thus an interference network with $\K$ cells, each consisting of a single Tx/Rx pair (i.e., a single mobile/BS pair). % Txs and  Rxs are equipped with $\L$ antennas and  
Networks have a regular interference pattern except at the network borders, with a focus on three different network topologies with short-range interference: 
\begin{itemize}
\item Wyner's linear symmetric model in Figure~\ref{fig5.10}(a), where Txs and Rxs are aligned on two parallel lines and interference is only from the two Txs on the left and the right of any given Tx/Rx pair. This topology models for example situations in a corridor or along a railway line or highway where BSs are aligned. Cooperation links are present between neighbouring Txs and between neighbouring Rxs. 
\item Wyner's hexagonal model in Figure~\ref{fig5.10}(b), where cells are assumed of hexagonal shape. Interference is from the six neighbouring cells. Cooperation links are present between BSs and between mobiles of neighbouring cells.
\item Sectorized hexagonal model in Figure~\ref{fig5.10}(c), where cells are again  of hexagonal shape. In this model, Txs and Rxs use directed antennas, allowing to divide each cell into three sectors with non-interfering communications, and interference is only from the neighbouring sectors in neighbouring cells, but not from sectors within the 
same cell. Here, a single mobile user is assumed in each sector, and thus three mobiles in each cell. Cooperation links are present between BSs of neighbouring cells and between mobiles in neighbouring sectors that are not in the same cell. 
\end{itemize}

Each Tx~$k \in [\K]$ wishes to convey a pair of independent  URLLC and eMBB messages  $M_k^{(\U)}$ and $M_k^{(\e)}$ to its corresponding Rx~$k \in [\K]$. URLLC Message $M_k^{(\U)}$ is of rate $R_k^{(\U)}$ and eMBB message $M_k^{(\e)}$ of rate $R_k^{(\e)}$. The focus is on the average URLLC and eMBB rates 
\begin{IEEEeqnarray}{rCl}
R^{(\U)} &:=&\frac{1}{\K} \sum_{k=1}^{\K}  R_k^{(\U)}\\
R^{(\e)} &:=&\frac{1}{\K} \sum_{k=1}^{\K}  R_k^{(\e)}.
\end{IEEEeqnarray}

 Consider a cooperation scenario where neighbouring Txs  cooperate during $\Dt>0$ rounds and neighbouring Rxs   during $\Dr>0$ rounds.
The total cooperation delay is constrained as 
\begin{equation}\label{eq:delay}
\Dt +\Dr \leq \D,
\end{equation} 
where $\D \ge 0$ is a given parameter of the system and the values of $\Dt$ and $\Dr$ are design parameters and can be chosen arbitrary such that \eqref{eq:delay} is satisfied. 
  During the $\Dt$ Tx-cooperation  rounds, each Tx can send  arbitrary  messages to its neighbours depending on the cooperation  messages it received in previous rounds and on its eMBB Message $M_k^{(\e)}$. In contrast, Tx-cooperation messages cannot depend on URLLC messages, as they are created only shortly before their transmission over the channel. The cooperative communication is assumed noise-free and the total cooperation load over all $\Dt$ Tx-cooperation rounds on each link is limited to $n \cdot \muT/2 \log(1+\P)$ bits. 
Each Tx forms then its channel inputs $X_k^n$ as a function of all its received cooperation messages $\vect{T}_k$ and both its URLLC message $M_k^{(\U)}$ and eMBB message $M_k^{(\e)}$:
\begin{IEEEeqnarray}{rCl}\label{xkn}
 X_k^n  =  {f}_k^{(n)} \Big( M_{k}^{(\U)}, M_{k}^{(\e)}, \vect{T}_k\Big).\IEEEeqnarraynumspace
\end{IEEEeqnarray}
Channel inputs at each Tx are subject to an average  block-power constraint $\P$. 

After receiving its channel outputs 
\begin{equation}\label{ykn}
 Y_{\tk}^n =   H_{k,k} \ X_k^n + \sum_{\hat k \in \mathcal I_{\tk}} {H}_{\hat k ,k}\   X_{\hat k}^n + Z_{k}^n,
\end{equation}
where ${Z}_{k}^n$ is  i.i.d. standard Gaussian noise, matrix $\mathsf{H}_{\hat k,k}$ models the channel from Tx~$\hat k$ to Rx $k$, which is assumed to be constant during the duration of communication and known by all terminals. 

Each Rx~$\tk$ immediately decodes its intended URLLC message:
\begin{equation}\label{mhatf}
\hat{{{M}}} _{\tk}^{(\U)} ={g_{\tk}^{(n)}}\big(  Y_{\tk}^{n}\big),
\end{equation} 
 using some decoding function $g_{\tk}^{(n)}$ on appropriate domains. %Throughout this section, we assume that the channel matrices $ \mathsf{H}_{\hat k ,k}$ are fixed  during the entire communication and known to every terminal. 
 Following this first decoding step, neighbouring Rxs communicate with each other during $\Dr$ Rx-cooperation rounds.  In each round, each Rx can send arbitrary messages to its neighbours that depend both on the previously received cooperation messages as well as on  its output signals. The cooperative communication is  assumed noise-free, but its total communication load over all $\Dr$ Rx-cooperation rounds on each link is restricted to $n \cdot \muR/2 \log(1+\P)$ bits. 
 At the end of these $\Dr$ Rx-cooperation rounds, each Rx~$\tk$ decodes its desired eMBB message as
\begin{equation}\label{mhats}
\hat{{M}}_{\tk}^{(\e)}={b_{\tk}^{(n)}}(  Y_{\tk}^n,\vect{Q}_{\tk}),
\end{equation}
where $\vect{Q}_{\tk}$ denotes all the Rx-cooperation messages received at Rx $\tk$ and $b_{\tk}^{(n)}$ is an appropriate decoding function. 

The focus of this section is on the \emph{Degrees of Freedom (DoF) region} of the described model, i.e., on the set of  possible pre-log factors $(\S^{(\U)}, \S^{(\e)})$ of  URLLC and eMBB rates that are simultaneously achievable in the limit of infinite powers $\P \to \infty$: 
\begin{IEEEeqnarray}{rCl}
\S^{(\U)} &:=& \lim_{\P \to \infty} \frac{ R^{(\U)}(\P)}{\frac{1}{2}\log \P} \\
\S^{(\e)} &:=& \lim_{\P \to \infty} \frac{ R^{(\e)}(\P)}{\frac{1}{2}\log \P},
\end{IEEEeqnarray}
where the pairs $(R^{(\U)}(\P), R^{(\e)}(\P))$ need to be simultaneously achievable for given power $\P$.

\input{Cell_association.tex}
 
\subsection{Coding Schemes}\label{sec:coding}
The following coding scheme was presented in \cite{HomaTCOM2021}. All Txs in the network are scheduled to either send their URLLC message, their eMBB message, or no message at all. The scheduled eMBB and URLLC messages are then jointly transmitted using an integrated URLLC/eMBB coding scheme. Different schedulings can be envisioned to achieve fairness and send all the required messages. Scheduling is described by  three sets $\mathcal T_{\text{silent}}$, $\mathcal T_{\U}$, and $\mathcal T_{\e}$, where
\begin{itemize}
\item Txs in $\mathcal T_{\text{silent}}$ are silenced and Rxs  in $\mathcal T_{\text{silent}}$ do not take any action. 
\item Txs in $\mathcal T_{\U}$ send only URLLC messages.  Tx/Rx pairs in $\mathcal{T}_{\U}$ are called \emph{URLLC Txs/Rxs}.
\item Txs in $\mathcal T_{\e}$ send only eMBB messages.  Tx/Rx pairs in $\mathcal{T}_{\e}$ are called  \emph{eMBB Txs/Rxs}.
\end{itemize}
Figure~\ref{fig5.10} illustrates the choices of the  $\mathcal T_{\text{silent}}$, $\mathcal T_{\U}$, and $\mathcal T_{\e}$  proposed in \cite{HomaTCOM2021} for Wyner's linear symmetric network, the hexagonal model, and the sectorized hexagonal model when the maximum number of allowed cooperation rounds is either $\D=6$ or $\D=8$. White colour is used for Tx/Rx pairs in $\mathcal T_{\text{silent}}$, yellow colour for pairs in $\mathcal T_{\U}$, and blue colour for pairs in $\mathcal T_{\e}$. 
The  set $\mathcal T_{\U}$ is chosen as large as possible but  so that URLLC transmissions
are interfered only by eMBB transmissions and not by other URLLC transmissions.

Consider the following joint coding scheme, which integrates both eMBB and URLLC messages. eMBB Txs describe quantized versions of their channel input signals  during the Tx-conferencing phase to their neighbouring URLLC Txs, which then precancel the interference
on their transmissions. URLLC Rxs can thus decode based on  interference-free channels.
After decoding, URLLC  Rxs   describe their decoded messages during
the Rx-conferencing phase to the adjacent eMBB Rxs, so as to  allow
them to pre-subtract the interference from URLLC messages
before decoding their intended eMBB messages. As a result, with the proposed scheduling and coding, 
URLLC messages can be decoded based on interference-free
outputs and  do not disturb the transmission of
eMBB messages. For the transmission of eMBB messages, either  CoMP transmission or CoMP reception is used, see Section~\ref{sec:CoMP}, but only on subnets. In fact, with the choice of    $\mathcal{T}_{\text{silent}}$  in Figure~\ref{fig5.10}, the networks decompose into small subnets so that  each subnet contains a master Tx/Rx that can be reached by any other Tx/Rx in the subnet with no more than $(\D-2)/2$ hops over the cooperation links. This ensures that CoMP transmission or reception in each subnet is possible with only $\D-2$ cooperation rounds. Since a single cooperation round is used to describe eMBB transmit signals to URLLC Txs and a single round is used to describe the decoded URLLC messages to eMBB Rxs, the scheme respects the maximum number $\D$ of total cooperation rounds. 

%To ensure fairness across all users and to send all URLLC messages and all eMBB messages, the described scheme can be time-shared for different, symmetric choices of the sets $\mathcal T_{\text{silent}}$, $\mathcal T_{\U}$, and $\mathcal T_{\e}$.
 
The coding scheme described above transmits both URLLC and eMBB messages. Variants thereof can be used to transmit only eMBB messages or only URLLC messages. More precisely,  since any eMBB message can also be treated as a URLLC message (this would mean impose stringent delay constraints also on some eMBB messages), the same scheme can also be used to  send only eMBB messages. An alternative for sending only eMBB messages,  is to silence again a set of Tx/Rx pairs and then directly employ CoMP transmission or CoMP reception on the set of non-silenced Txs/Rxs. Both schemes achieve the same DoF, but depending on the specific network they require  larger or smaller cooperation rates $\muT$ and $\muR$. 

A simple way to send only URLLC messages  is to choose a largest possible set of non-interfering Tx/Rx pairs and to silence all other Tx/Rx pairs. For Wyner's linear symmetric network this is optimal. For the two hexagonal models, and for certain channel coefficients, better performance is possible using the  interference alignment techniques in \cite{Jafar}.

\subsection{Results on the Joint eMBB/URLLC DoF region}\label{sec:results}
This subsection presents the achievable  eMBB/URLLC DoF region achieved by the schemes in the previous  subsection on the three network topologies in Figure~\ref{fig5.10}, and compares them  to the outer bounds derived in \cite{HomaTCOM2021}.

First consider Wyner's linear  symmetric network. For this network and for sufficiently large cooperation prelog factors $\muT$ and $\muR$,   all DoF pairs $(\S^{(\U)}, \S^{(\e)})$ in the  DoF region are  achieved by the schemes  described in the previous subsection or by  time-sharing different versions thereof. The DoF region is given by the set of all DoF pairs $(\S^{(\U)}, \S^{(\e)})$ that satisfy % allowing for the probability of error tending to 0 as the blocklength $n\to \infty$ is given by the pairs satisfying
\begin{subequations}\label{eq:Wyner}
\begin{IEEEeqnarray}{rCl}
0 \leq \S^{(\U)} &\le& \frac{1}{2},\\
0\leq \S^{(\U)} + \S^{(\e)} &\le&\frac{\D+1}{\D+2}.
\end{IEEEeqnarray}
\end{subequations}
One notices that the sum-DoF of the system is limited by the maximum number of allowed cooperation rounds $\D$. Moreover, the stringent delay constraint on URLLC messages does not penalize the maximum achievable sum-DoF, which is equal to $\L \cdot \frac{\D+1}{\D+2}$, irrespective of $\S^{(\U)}$. 

 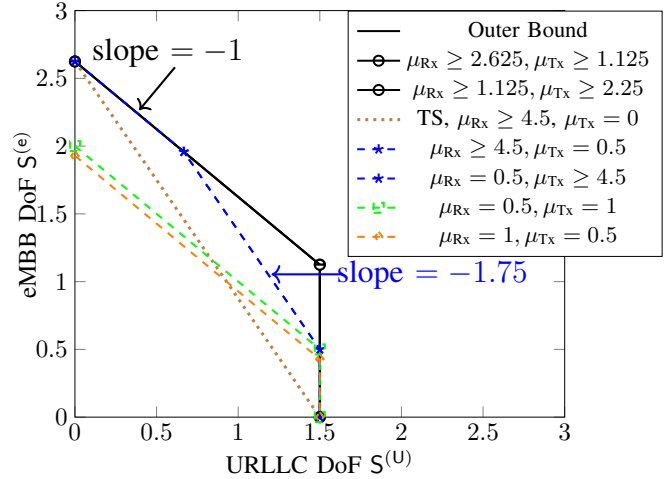
\begin{figure}[t]	
 	\centering
 	\begin{tikzpicture}[scale=0.95]
 	\begin{axis}[
 	xlabel={\small {URLLC DoF $\S^{(\U)}$ }},
 	ylabel={\small {eMBB DoF $\S^{(\e)}$ }},
 	xlabel style={yshift=.5em},
 	ylabel style={yshift=-1.25em},
 	xmin=0, xmax=3,
 	ymin=0, ymax=3,
 	xtick={0,0.5,1,1.5,2,2.5,3, 3.5},
 	ytick={0,0.5,1,1.5,2,2.5,3},
 	yticklabel style = {font=\small,xshift=0.25ex},
 	xticklabel style = {font=\small,yshift=0.25ex},
 	 legend style={at={(1.2,1)}},
 	]
 	\addplot[color=black,thick]coordinates {(0,2.625)(1.5,1.125) (1.5,0)};	
 	\addplot[color=black,mark=halfcircle,thick]coordinates {(0,2.625)(1.5,1.125) (1.5,0)};
 	\addplot[color=black,mark=halfcircle,thick]coordinates {(0,2.625)(1.5,1.125) (1.5,0)};
 	\addplot[color=brown, very thick,dotted]coordinates {(0,2.625)(1.5,0)};
 	\addplot[color=blue,mark=star,thick, dashed]coordinates{(0,2.625)(0.6666,1.9583)(1.5,0.5) (1.5,0)};
 	\addplot[color=blue,mark=star,thick, dashed]coordinates{(0,2.625)(0.6666,1.9583)(1.5,0.5) (1.5,0)};

	\addplot[color=green,mark=square, thick,dashed]coordinates{(0,2.0)(1.5,0.5) (1.5,0)};
\addplot[color=orange,mark=diamond,thick, dashed]coordinates{(0, 1.928 )(1.5,0.428) (1.5,0)};

 	\legend{{\footnotesize Outer Bound}, {\footnotesize $ \muR\ge 2.625, \muT\ge 1.125$},{\footnotesize $ \muR\ge 1.125, \muT\ge 2.25$},{\footnotesize TS, $ \muR\ge 4.5$, $\muT= 0$},{\footnotesize $\muR \ge 4.5, \muT =0.5$},  {\footnotesize $\muR = 0.5, \muT \ge 4.5$},{\footnotesize $\muR = 0.5, \muT =1$},{\footnotesize $\muR = 1, \muT =0.5$}}  
 	
 	\end{axis}
 	
 	\node[draw =none] (s2) at (1.3,5.1) { slope $=-1$};
 	\draw [->,  thick ] (1.4,4.9)--(0.9,4.3);
 \node[draw =none, blue] (s2) at (4.5+0.5,2 ) { slope $=-1.75$};
 	\draw [->, blue, thick ] (3.25+0.5,2)--(2.25+0.5,2);
 	\end{tikzpicture}
 	%\vspace{-0.5cm}
 	\caption{Bounds on DoF region for Wyner's symmetric model for  different values of $\muR$ and $\muT$, and  $\D=6$. The brown dotted line represents the pure scheduling performance. }
 	\label{fig5a}
 	%\vspace{-0.5cm}
 	\end{figure}

 \begin{figure*}[b]
 	\centering
 	\begin{tikzpicture}[scale=0.95]
 	
 	\begin{axis}[
 	xlabel={\small {$\S^{(\U)}$ }},
 	ylabel={\small {$\S^{(\e)}$ }},
 	xlabel style={yshift=.5em},
 	ylabel style={yshift=-1.25em},
 	xmin=0, xmax=1.7,
 	ymin=0, ymax=3,
 	xtick={0,0.5,1,1.5,2},
 	ytick={0,0.5,1,1.5,2, 2.5,3},
 	yticklabel style = {font=\small,xshift=0.25ex},
 	xticklabel style = {font=\small,yshift=0.25ex},
	legend style={at={(1.65,1)}},
 	]
 	\addplot[color=black,thick]coordinates {(0,2.83)(1.5,1.33) (1.5,0)};	
 	\addplot[color=black,mark=halfcircle,thick]coordinates {(0,2.4400)(0.8125,1.5) (1,0)};
 	\addplot[color=black,mark=halfcircle,thick]coordinates {(0,2.4400)(0.8125,1.5) (1,0)};
 	\addplot[color=brown,dotted,very thick]coordinates {(0,2.4400)(1,0)};
 	
 	\addplot[color=blue,mark=star,thick, dashed]coordinates {(0,2.3125)(0.8125,1.5) (1,0)};
 	\addplot[color=blue,mark=star,thick, dashed]coordinates {(0,2.3125)(0.8125,1.5) (1,0)};
 	
	\addplot[color=green,mark=square, thick,dashed]coordinates{(0,2.4400)(0.13,2.2896)(0.9700, 0.24) (1,0)};
	\addplot[color=green,mark=square, thick,dashed]coordinates{(0,2.4400)(0.13,2.2896)(0.9700, 0.24) (1,0)};%	
	\addplot[color=orange,mark=diamond,thick, dashed]coordinates{(0,1.6)(0.88,0.96) (1,0)};
	\addplot[color=red, thick, dashed]coordinates{(0,1.6)(0.8928,0.8571) (1,0)};

\small {
 	\legend{{\small Outer Bound}, {\small $ \muR\ge2.4, \muT\ge0.6$},{ $ \muR\ge0.63, \muT\ge2.4$}, { TS,  $ \muR\ge 2.4 \muT = 0$},  { $1.7\le\muR <2.4 , \muT \ge 0.6$},{ $\muR \ge 0.6, 1.5 \le \muT < 2.4 $}, {$\muR\ge 2.4, \muT =0.1$},{$\muR = 0.1, \muT \ge 2.4$},{$\muR = 0.5, \muT =1$},{$\muR = 1, \muT =0.5$}} } 
 	
 	\end{axis}

 	 \node[draw =none, rotate = -20] (s2) at (1.9,5.2-0.7 -0.2-0.1-0.1) { \small slope $=-1.15$};
 	 \draw [->, thick] (1.85,5.1-0.7-0.3-0.1)--(1.85-0.1,4.3-0.65);
 	 
 	 \node[draw =none, brown] (s2) at (1.2,1) { \small slope $=-2.44$};
 	 \draw [->, thick, brown] (2.5,1)--(3.1,1);
 	\end{tikzpicture}
 	
 	\caption{Inner and outer  bounds on the DoF region for the hexagonal model  for $\D = 8$ and different values of $\muR$ and $\muT$. The brown dotted  line shows the pure time-sharing region. }
 	\label{fig5.6}
 \end{figure*}
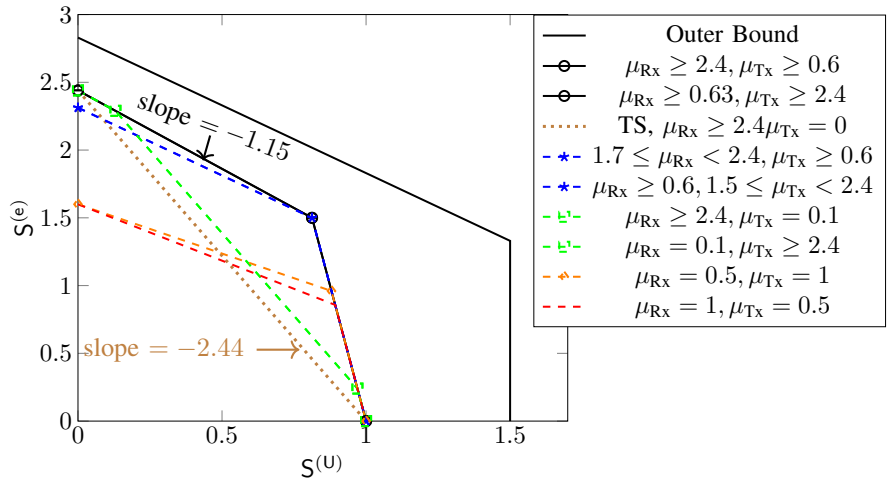

For smaller cooperation prelog factors $\muT,\muR$ this is not the case, as can be seen at hand of Figure~\ref{fig5a}, which shows inner and outer bounds on the DoF region derived in \cite{HomaTCOM2021}. The inner bound is achieved by time-sharing the coding schemes in Subsection~\ref{sec:coding}, and significantly improves over a pure scheduling approach that time-shares between a system sending only URLLC messages or  only eMBB messages. We notice that for  $\muR\ge 2.625$ and $\muT \ge 1.125$ the inner and outer bounds match. The inner bound  is achieved by the schemes in Section~\ref{sec:coding} employing CoMP reception for eMBB messages. For  $\muR \ge 1.125$ and $\muT\ge 2.25$ the inner and outer bounds also match, and are achieved by the same schemes, but employing CoMP transmission. %In this network, CoMP transmission thus requires a smaller total cooperation rate than CoMP reception.  
When  only one of the two cooperation prelogs $\muT$ or $\muR$ is large and the other small  (e.g., $\muR \ge 4.5$ and $\muT =0.5$; or $\muT \ge 4.5$ and $\muR = 0.5$) the inner bound matches the outer bound only for $\S^{(\U)}$ below a given threshold. For URLLC DoFs  $\S^{(\U)}$ exceeding this threshold,  the maximum eMBB DoF $\S^{(\e)}$ achieved by the schemes in Subsection~\ref{sec:coding} decreases linearly  with $\S^{(\U)}$. For example, for $\D = 6$ and  $(\muR \ge 4.5, \muT =0.5)$ or $(\muT \ge 4.5,\muR = 0.5)$, beyond this threshold, when one increases the URLLC DoF  $\S^{(\U)}$  by $\Delta$,  then the  eMBB Dof $\S^{(\e)}$ decreases by  approximately  $1.75\Delta$ and  the sum DoF       by $0.75 \Delta$. Yet another behavior is observed when both $\muR$ and $\muT$ are moderate or small, e.g., $\muR = 0.5$ and $\muT = 1$ or $\muR = 1$ and $\muT = 0.5$. In this case, the sum DoF achieved by the inner bound is not at its maximum value, but constant over all  regimes of $\S^{(\U)}$. The overall performance of the system is thus again not limited by the stringent delay constraints on URLLC messages, but simply by the available cooperation rates.  %Moreover,  the inner bounds remain unchanged for $\D=6, 8,10$ because even for $\D>6$ it is more advantageous to reduce the number of cooperation rounds  to $6$ in order to satisfy the cooperation prelogs  than to time-share different schemes with $\D$ cooperation rounds. %The brown dotted line is the  result of time-sharing the scheme in Subsection \ref{sub:symslow} that transmit  only eMBB messages  with the scheme in Subsection \ref{sub:symfast} when only URLLC messages are transmitted. %The scheme in Subsection \ref{sub:symslow} is based on CoMP reception and  to achieve the maximum ``slow'' MG requires  $\muR \ge 4.5$ for $\D = 6$ and $\muR  \ge 7.5$ for $\D = 10$.   The sum-MG in this scheme decreases linearly with the ``fast'' MG.

 Figure~\ref{fig5.6} shows the inner and outer  bounds on the DoF region proposed in \cite{HomaTCOM2021} for the hexagonal model  when $\D=8$ and for different values of  $\muR$ and $\muT$. Unlike in Wyner's symmetric model, the sum DoF achieved by the schemes in Subsection~\ref{sec:coding} always decreases as $\S^{(\U)}$ increases, irrespective of the cooperation prelogs $\muT,\muR$. Moreover, the  maximum $\S^{(\U)}=\frac{\L}{3}$  is only achieved for    $\S^{(\e)}=0$. %We remark here that for certain channel matrices  (in fact for many but not for all) ``fast" MG $\S^{(\e)}$ is achievable using interference alignment \cite{Etkin,Motahari, Stotz}. For these channel matrices of course our inner bound can be improved accordingly.	

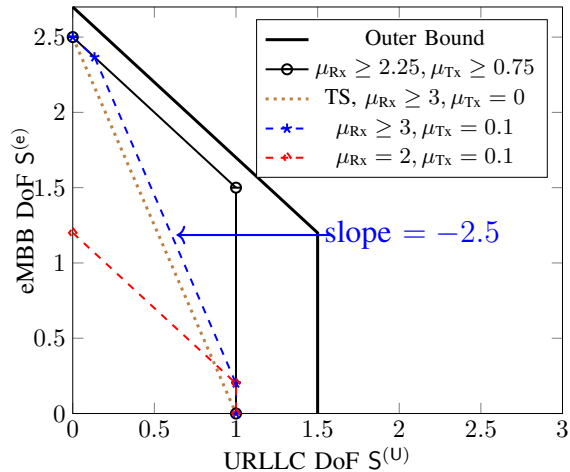
\begin{figure}[t]
 	\centering
 	\begin{tikzpicture}[scale=0.95]
 	
 	\begin{axis}[
 	xlabel={\small {URLLC DoF $\S^{(\U)}$ }},
 	ylabel={\small {eMBB DoF $\S^{(\e)}$ }},
 	xlabel style={yshift=.5em},
 	ylabel style={yshift=-1.25em},
 	xmin=0, xmax=3,
 	ymin=0, ymax=2.7,
 	xtick={0,0.5,1,1.5,2,2.5,3},
 	ytick={0,0.5,1,1.5,2,2.5},
 	yticklabel style = {font=\small,xshift=0.25ex},
 	xticklabel style = {font=\small,yshift=0.25ex},
 	 legend pos=north east,
 	]
 	\addplot[color=black,very thick]coordinates {(0,2.7000)(1.5,1.2) (1.5,0)};	
 	\addplot[color=black,mark=halfcircle,thick]coordinates {(0,2.5000)(1,1.5) (1,0)};
 	\addplot[color=brown ,dotted,very thick]coordinates {(0,2.5000)(1,0)};
 	
	\addplot[color=blue,mark=star, thick,dashed]coordinates{(0,2.5000)(0.1333,2.3667)(1, 0.2) (1,0)};

                      \addplot[color=red,mark=diamond,thick, dashed]coordinates {(0,1.2)(1,0.2) (1,0)};

 	\legend{{\footnotesize Outer Bound}, {\footnotesize $ \muR\ge2.25, \muT\ge 0.75$},{\footnotesize TS, $ \muR\ge3, \muT = 0$}, {\footnotesize $ \muR\ge 3, \muT=0.1$},  {\footnotesize $\muR = 2, \muT =0.1$}}  
 	
 	\end{axis}

 	 \node[draw =none, blue] (s2) at (4.75,2.5 ) { slope $=-2.5$};
 	\draw [->, blue, thick ] (3.6,2.5)--(1.45,2.5);
 	\end{tikzpicture}
 	
 	\caption{Inner and outer bounds on the DoF region  for the sectorized hexagonal model  for $\D = 4$ and different values of $\muR$ and $\muT$. The brown dotted line indicates the pure scheduling performance.}
 	\label{fig5.11}
 \end{figure}

Figure~\ref{fig5.11} shows inner and outer bounds on the DoF region for the sectorized hexagonal model when $\D=4$. We notice that  when both $\muR$ and $\muT$ are above given thresholds, $(\muT \geq 0.75, \muR\geq 2.25)$, then the combined scheme integrating both URLLC and eMBB messages in Subsection~\ref{sec:coding} simultaneously achieves maximum URLLC DoF and maximum sum-DoF.   If only  one of the two cooperation prelogs is  very high  but the other one small, the scheme achieves maximum sum-DoF only for small  URLLC DoFs.  The reason is that  the integrated scheme in Subsection~\ref{sec:coding} that jointly sends   URLLC \emph{and} eMBB messages inherently requires both Tx- \emph{and} Rx-cooperation of sufficiently high cooperation prelogs, whereas Tx- \emph{or} Rx-cooperation are sufficient for the  scheme that  sends only eMBB messages. 
% As a consequence, the maximum  $\S^{(\e)}$ that our schemes achieve for large $\S^{(\U)}$ highly depends on the smaller of the two cooperation prelogs $\muT$ and $\muR$. 
% 3) When both $\muT,\muR$ are moderate,  we can still achieve the MG pair  $(\S_{\text{both}}^{(\U)}, \S_{\text{both}}^{(\U)})$ but not $(0, \S^{(\e)}_{\max})$. In the regime of small $\S^{(\U)}$ there is thus a penalty in $\S^{(\e)}$ and sum MG compared to the case of high cooperation prelogs but not in the regime of large $\S^{(\U)}$. 4) Finally, when both cooperation prelogs become small then neither of the two points  $(0, \S^{(\e)}_{\max})$ and $(\S_{\text{both}}^{(\U)}, \S_{\text{both}}^{(\U)})$  is achievable anymore. 
 
%The brown dotted line is the resulting region under the traditional scheduling scheme that time-shares the scheme achieving the point $(0, \S^{(\e)}_{\max})$ with the scheme achieving the point $(\S_{\text{no-coop}}, 0)$. To achieve the point  $(0, \S^{(\e)}_{\max})$, this scheme requires $\muR \ge 2.4$ and $\muT = 0$ while using  CoMP reception, and $\muR = 0$ and $\muT \ge 2.4$ while using CoMP transmission. Comparing the slope of this line with the slopes of the regions achieved under our proposed scheme show that the penalty  on the sum-MG caused by the transmission of URLLC  messages is large in this scheme. 

%Figure~\ref{fig5.11} illustrates  inner and outer  bounds on the DoF  region for the sectorized model when $\D=4$, and for different values of  $\muR$ and $\muT$. When $\muR \ge 2.25$ and $\muT\ge 0.75$, there is no penalty in sum DoF even at maximum URLLC DoF. It also can be seen from this figure that transmitting ``fast'' messages using the traditional  time-sharing scheme (brown dotted line) at any ``fast'' MG has a penalty on sum-MG. } 

To summarize, for all three considered network models  the joint scheme in Subsection~\ref{sec:coding} that integrates both URLLC and eMBB messages achieves maximum  sum-DoF at high (or maximum) URLLC DoFs whenever the cooperation rates are sufficiently large. In this case, the stringent delay constraints on the URLLC messages do not harm the overall system performance. For smaller cooperation rates either the maximum sum-DoF is decreased or it is the same as with high cooperation prelogs but can only be achieved for small URLLC DoFs. 

The described integrated coding scheme inherently requires at least a single cooperation round both at the Tx-side as well as at the Rx-side. The work in \cite{HomaEntropy}   also considered a scenario with only Rx- or only Tx-cooperation. It was shown that when only Rxs  or only Txs can cooperate, then the ideal performance in  \eqref{eq:Wyner} is not possible. Instead for sufficiently large cooperation rates the DoF region is given by the set of all rate-pairs $(\S^{(\U)}, \S^{(\e)})$ satisfying \cite{HomaEntropy}
\begin{subequations}\label{eq:WynerRx}
\begin{IEEEeqnarray}{rCl}
0 \leq 2 \S^{(\U)} + \S^{(\e)}  &\le&1,\\
0\leq  \S^{(\U)} + \S^{(\e)} &\le&\frac{\D+1}{\D+2}.
\end{IEEEeqnarray}
\end{subequations}
The  maximum sum-DoF is thus not decreased compared to a scenario with Tx- \emph{and} Rx-cooperation. However, this maximum sum-DoF is only achievable for URLLC DoF $\S^{(\U)} \leq  \frac{1}{\D+2}$. We conclude that the stringent delay constraint inherently limits the overall system performance for moderate or large URLLC DoFs when only Rxs can cooperate. In fact, in this regime, increasing the URLLC DoF by $\Delta$ requires decreasing the eMBB DoF by $2\Delta$ and the sum-DoF by $\Delta$. Similar conclusions also hold for smaller cooperation prelogs and even in the non-asymptotic regime of finite powers \cite{HomaEntropy}. 

\subsection{Random User Activities} \label{sec:random}
In practical systems, URLLC messages (and sometimes even eMBB messages) arrive in a random and bursty fashion and consequently in any given block, some Txs do not have  an URLLC message to transmit.
We consider the \emph{random user-activity and random arrival model} proposed in \cite{HomaITW2021}, where each Tx is active with probability $\rho$, independent of all other Txs. If a Tx
is active,  it sends an eMBB message to its corresponding Rx, and moreover, with probability $\rho_f$,  it also sends an additional URLLC message. Both the activity and arrival realizations are assumed to be known to all  terminals in the network. 

The DoF of \emph{all} URLLC messages in the system is fixed  and given by  $\S^{(\U)}$, whereas the  \emph{eMBB} DoF can vary over the variuos eMBB messages, and the quantity of interest is  the \emph{expected  average}  DoF $\S^{(\e)}$ over all eMBB messages. (Similarly to the BC scenario in Section~\ref{sec:BC}, the expected average eMBB DoF  accounts for the possibility that eMBB messages are   sent over multiple URLLC arrival blocks.) The same random-user activity model (but without mixed delays and random message arrivals) was already  considered in \cite{Somekh2008,Levy2008,Somekh2009}, where it was observed that under this model the networks considered in Figure~\ref{fig5.10} decomposes into non-interfering subnets.

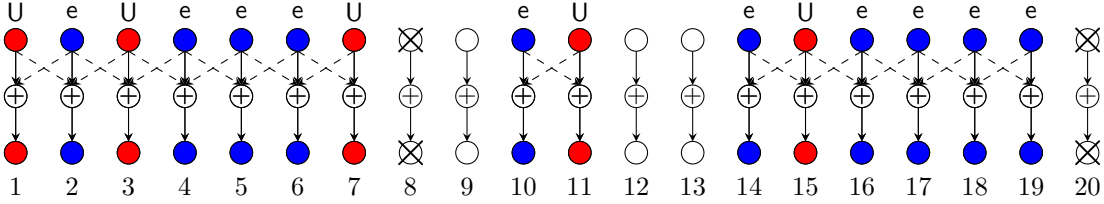
\begin{figure*}[t]
%\footnotesize
  \centering
%\begin{subfigure}{1\textwidth}
%\centering
\begin{tikzpicture}[scale=1.5, >=stealth]
\centering
\tikzstyle{every node}=[draw,shape=circle, node distance=0.5cm];
 \foreach \j in {1,...,20} {
 \draw (-3.5 + 0.5*\j, 1.5) circle (0.1);
\node[draw =none] (s2) at (-3.5+ 0.5*\j,1 ) {\footnotesize$+$};
\draw (-3.5 +0.5*\j, 1) circle (0.1);
 \draw (-3.5 + 0.5*\j, 0.5) circle (0.1);
 \draw   [->] (-3.5+ 0.5*\j,1.9-0.5)-- (-3.5+ 0.5*\j,1.1);
 \draw   [->] (-3.5+ 0.5*\j,0.9)-- (-3.5+ 0.5*\j,0.6);
 }
  \foreach \j in {1,...,6,10,14,15,16,17,18} {
  \draw   [->, dashed] (-3.5+ 0.5*\j,1.9-0.5)-- (-3.5+ 0.5*\j + 0.5,1.1);
  }
    \foreach \j in {2,...,7,11,15,16,17,18,19} {
  \draw   [->, dashed] (-3.5+ 0.5*\j,1.9-0.5)-- (-3.5+0.5*\j - 0.5,1.1);
  %\draw   [->, dashed] (-3.5+ 0.5*\j,1.9)-- (-3.5+ 0.5*\j - 0.5,1.1);
 }
  \foreach \j in {1,3,7,11,15} {
 \draw [fill = red](-3.5 + 0.5*\j, 1.5) circle (0.1);
\node[draw =none] (s2) at (-3.5+ 0.5*\j,1 ) {\footnotesize$+$};
\draw (-3.5 +0.5*\j, 1) circle (0.1);
 \draw [fill = red] (-3.5 + 0.5*\j, 0.5) circle (0.1);
 \draw   [->] (-3.5+ 0.5*\j,1.9-0.5)-- (-3.5+ 0.5*\j,1.1);
 \draw   [->] (-3.5+ 0.5*\j,0.9)-- (-3.5+ 0.5*\j,0.6);
%  \draw   [->, dashed] (-3.5+ 0.5*\j,1.9-0.5)-- (-3.5+ 0.5*\j + 0.5,1.1);
 % \draw   [->, dashed] (-3.5+ 0.5*\j,1.9-0.5)-- (-3.5+0.5*\j - 0.5,1.1);
  %\draw   [->, dashed] (-3.5+ 0.5*\j,1.9)-- (-3.5+ 0.5*\j - 0.5,1.1);
 }
   \foreach \j in {2,4,5,6,10,14,16,17, 18,19} {
 \draw [fill = blue](-3.5 + 0.5*\j, 1.5) circle (0.1);
\node[draw =none] (s2) at (-3.5+ 0.5*\j,1 ) {\footnotesize$+$};
\draw (-3.5 +0.5*\j, 1) circle (0.1);
 \draw [fill = blue] (-3.5 + 0.5*\j, 0.5) circle (0.1);
 \draw   [->] (-3.5+ 0.5*\j,1.9-0.5)-- (-3.5+ 0.5*\j,1.1);
 \draw   [->] (-3.5+ 0.5*\j,0.9)-- (-3.5+ 0.5*\j,0.6);
%  \draw   [->, dashed] (-3.5+ 0.5*\j,1.9-0.5)-- (-3.5+ 0.5*\j + 0.5,1.1);
%  \draw   [->, dashed] (-3.5+ 0.5*\j,1.9-0.5)-- (-3.5+0.5*\j - 0.5,1.1);
  %\draw   [->, dashed] (-3.5+ 0.5*\j,1.9)-- (-3.5+ 0.5*\j - 0.5,1.1);
 }
\small
 \node[draw =none] (s2) at (-3.5+0.5,0.2) {$1$};
\node[draw =none] (s2) at (-3.5+1,0.2) {$2$};
\node[draw =none] (s2) at (-3.5+1.5,0.2) {$3$};
\node[draw =none] (s2) at (-3.5+2,0.2) {$4$};
\node[draw =none] (s2) at (-3.5+2.5,0.2) {$5$};
\node[draw =none] (s2) at (-3.5+3,0.2) {$6$};
\node[draw =none] (s2) at (-3.5+3.5,0.2) {$7$};
\node[draw =none] (s2) at (-3.5+4,0.2) {$8$};
\node[draw =none] (s2) at (-3.5+4.5,0.2) {$9$};
\node[draw =none] (s2) at (-3.5+5,0.2) {$10$};
\node[draw =none] (s2) at (-3.5+5.5,0.2) {$11$};
\node[draw =none] (s2) at (-3.5+6,0.2) {$12$};
\node[draw =none] (s2) at (-3.5+6.5,0.2) {$13$};
\node[draw =none] (s2) at (-3.5+7,0.2) {$14$};
\node[draw =none] (s2) at (-3.5+7.5,0.2) {$15$};
\node[draw =none] (s2) at (-3.5+8,0.2) {$16$};
\node[draw =none] (s2) at (-3.5+8.5,0.2) {$17$};
\node[draw =none] (s2) at (-3.5+9,0.2) {$18$};
\node[draw =none] (s2) at (-3.5+9.5,0.2) {$19$};
\node[draw =none] (s2) at (-3.5+10,0.2) {$20$};

 %\node[draw =none] (s2) at (-3.5+0.5,0) {$\hat M_1^{(F)}$};\U
%___________________________________\e
\node[draw =none] (s2) at (-3.5+0.5,1.75) {$\U$};
\node[draw =none] (s2) at (-3.5+1,0.2+1.55) {$\e$};
\node[draw =none] (s2) at (-3.5+1.5,0.2+1.55) {$\U$};
\node[draw =none] (s2) at (-3.5+2,0.2+1.55) {$\e$};
\node[draw =none] (s2) at (-3.5+2.5,0.2+1.55) {$\e$};
\node[draw =none] (s2) at (-3.5+3,0.2+1.55) {$\e$};
\node[draw =none] (s2) at (-3.5+3.5,0.2+1.55) {$\U$};
%\node[draw =none] (s2) at (-3.5+4,0.2+1.55) {$\e$};
%\node[draw =none] (s2) at (-3.5+4.5,0.2+1.55) {$9$};
\node[draw =none] (s2) at (-3.5+5,0.2+1.55) {$\e$};
\node[draw =none] (s2) at (-3.5+5.5,0.2+1.55) {$\U$};
%\node[draw =none] (s2) at (-3.5+6,0.2+1.55) {$12$};
%\node[draw =none] (s2) at (-3.5+6.5,0.2+1.55) {$\U$};
\node[draw =none] (s2) at (-3.5+7,0.2+1.55) {$\e$};
\node[draw =none] (s2) at (-3.5+7.5,0.2+1.55) {$\U$};
\node[draw =none] (s2) at (-3.5+8,0.2+1.55) {$\e$};
\node[draw =none] (s2) at (-3.5+8.5,0.2+1.55) {$\e$};
\node[draw =none] (s2) at (-3.5+9,0.2+1.55) {$\e$};
\node[draw =none] (s2) at (-3.5+9.5,0.2+1.55) {$\e$};
%\node[draw =none] (s2) at (-3.5+10,0.2+1.55) {$\e$};

 \node[draw =none, rotate =45] (s2) at ( 0.5,1.5) {\huge$+$};
  \node[draw =none, rotate =45] (s2) at ( 0.5,0.5) {\huge$+$};
  \node[draw =none, rotate =45] (s2) at ( 6.5,1.5) {\huge$+$};
  \node[draw =none, rotate =45] (s2) at (6.5,0.5) {\huge$+$};
%  \foreach \i in {1,3,15} {
%  \draw [<-, very thick, blue ] (-3.5 + 0.5*\i+0.1, 1.5) -- (-3.5 + 0.5*\i + 0.5- 0.1, 1.5);
%  }
%   \foreach \i in {2,6,10,14,18} {
%  \draw [->, very thick, blue ] (-3.5 + 0.5*\i+0.1, 1.5) -- (-3.5 + 0.5*\i + 0.5- 0.1, 1.5);
%  }
%  
%   \foreach \i in {2,6,10,14} {
%  \draw [<-, very thick, red ] (-3.5 + 0.5*\i+0.1, 0.5) -- (-3.5 + 0.5*\i + 0.5- 0.1, 0.5);
%  }
%  
%    \foreach \i in {1,15} {
%  \draw [->, very thick, red ] (-3.5 + 0.5*\i+0.1, 0.5) -- (-3.5 + 0.5*\i + 0.5- 0.1, 0.5);
%  }
%     \foreach \i in {18} {
%  \draw [<-, very thick, blue ] (-3.5 + 0.5*\i+0.1, 0.5) -- (-3.5 + 0.5*\i + 0.5- 0.1, 0.5);
%  }
%   \foreach \j in {4,16} {
% \draw [very thick, green](-3.5 + 0.5*\j, 0.5) circle (0.12);
%  }
\end{tikzpicture}

  \caption{Wyner's symmetric linear network with random user activity and random arrival. $\D = 6$. }
   \label{fig-exp1}

  \end{figure*}

The same decomposition happens in the mixed-delay and random message arrivals model. As a consequence, an independent instance of the schemes in Subsection~\ref{sec:coding} should be applied to each subnet, where  the schemes  however have to be further adapted to the random URLLC message arrival situation. In particular, the scheduling (choices of sets $\mathcal{T}_{\textnormal{silent}}, \mathcal{T}_{\U}, \mathcal{T}_{\e}$) needs to be adapted to the actual URLLC messages present in a subnet. The work in \cite{HomaITW2021} proposes such a new scheduling approach, which on Wyner's linear symmetric network time-shares between a scheduling that sends URLLC messages at odd Txs and a second scheduling that sends URLLC messages at even Txs. (This allows to achieve a symmetric URLLC DoF over all users having a URLLC message to send.) The scheduling for odd URLLC Txs is illustrated in  Figure~\ref{fig-exp1} for Wyner's linear symmetric model and specific realizations of the user activities and URLLC message arrivals. In the presented  example,  Txs 9, 12, 13 are inactive and Txs 1, 3, 7, 11, and 15 have an URLLC message to send. The network thus decomposes into three subnets: the first includes Tx/Rx pairs 1--8, the second  includes Tx/Rx pairs 10--11, and the third includes Tx/Rx pairs 14--20.  In  each subnet, an independent instance of  the integrated scheme of Subsection~\ref{sec:coding} is applied,  but where the eMBB message at Tx 19 is  treated as  URLLC messages to comply with the scheme.  In particular, Txs 8 and 20 are silenced because the maximum allowed cooperation delay equals $\D=8$, and Txs/Rxs 4 and 16 act as master Txs/Rxs  in the CoMP scheme.%The blue arrows  in Figure~\ref{fig-exp1} indicate 

  %\begin{itemize} 
%  \item When $\rho = 1$, the fundamental MG region $\S^*(\rho, \rho_f)$ includes all nonnegative pairs $(\S^{(\U)}, \; \S^{(\e)})$ satisfying
% \begin{subequations}
% \begin{IEEEeqnarray}{rCl}
% \S^{(\U)} &\le& \frac{\rho_f}{2} \\
% \S^{(\e)} +\S^{(\U)} & \le & 1
% \end{IEEEeqnarray}
% \end{subequations}
For sufficiently large cooperation rates, the approach in \cite{HomaITW2021} achieves all DoF pairs  $(\S^{(\U)}, \; \S^{(\e)})$ satisfying
 \begin{IEEEeqnarray}{rCl}
 \S^{(\U)} &\le& \frac{\rho \rho_f}{2}, \label{eq:a} \\
 \S^{(\e)} + M \cdot  \S^{(\U)} & \le & \rho - \frac{(1-\rho) \rho^{\D+2}}{1-\rho^{\D+2}},\label{eq:b}
 \end{IEEEeqnarray}
 where 
 \begin{IEEEeqnarray}{rCl} \label{eq:M}
 M \triangleq 1+ \frac{(1-\rho)^2 \rho^{\D+2}}{\rho \rho_f(1-\rho^{\D+2})} +\frac{(1-\rho)^2  \rho^{\D+1}  (1-\rho_f)^{\frac{\D}{2}}}{\rho \rho_f (1- \rho^{\D+2}(1-\rho_f)^{\frac{\D}{2}+1})}. \IEEEeqnarraynumspace
 \end{IEEEeqnarray}
% and $\Delta$  equals: 
% \begin{IEEEeqnarray}{rCl} \label{eq:delta}
%\Delta& \triangleq& \frac{(1+\rho)  (1- \rho_f)^{\frac{\D}{2}+1}}{2 \left(1- \rho^{\D+2} (1-\rho_f)^{\frac{\D}{2}+1}\right)} \\ \notag 
%&& - \frac{\left (\rho(1-\rho_f) \right )^{\D +1}(1 - \left ( (1-\rho_f)\rho^2 \right )^{\frac{\D}{2}}) }{2(\left(\rho^{2}(1-\rho_f)\right )^{(\D+1)}-1)^2} \\ \notag
%&&  \quad  \times \left ((1+ \rho) \rho^{\D+1} (1- \rho_f)^{\frac{\D}{2} +1} + 2 (1+\rho(1-\rho_f))  \right)
% \end{IEEEeqnarray}
By means of an information-theoretic converse, it can be shown that all DoF pairs $(\S^{(\U)}, \; \S^{(\e)})$ not
satisfying \eqref{eq:a} and %the following two conditions:
 \begin{IEEEeqnarray}{rCl}
% \S^{(\U)} &\le& \frac{\rho \rho_f}{2}, \label{eq:conv1}\\
 \S^{(\e)} + \S^{(\U)} & \le & \rho - \frac{(1-\rho) \rho^{\D+2}}{1-\rho^{\D+2}} \label{eq:conv2}
 \end{IEEEeqnarray}
 cannot lie in the DoF region. Constraints \eqref{eq:a} and \eqref{eq:conv2} thus provide an outer bound on the DoF region. 
Notice that this outer bound and the   inner bound given by \eqref{eq:a} and \eqref{eq:b} only differ in the factors, $M>1$ or $1$, preceding the URLLC DoF $\S^{(\U)}$ in the bounds \eqref{eq:b} and \eqref{eq:conv2}, respectively. These factors   are   close whenever $\D \geq 10$, and as a consequence also the presented inner and outer bounds are close. For small values of $\rho$ the factors are already close for $\D\geq 4$. Moreover, for small values of $\rho$ and $\D\geq 4$, the right-hand sides of \eqref{eq:b} and \eqref{eq:conv2}, are approximately equal to $\rho$, irrespective  of $\D$. This indicates that for small values of $\rho$, increasing the number of cooperation rounds $\D$ beyond 4 (and thus further increasing the  delay of eMBB messages)  does not improve the DoF region of the system. The reason behind this phenomenon is that a large number of cooperation rounds $\D$ is only useful in subnets with a large number of consecutive active Txs, and such subnets are very rare when  the random user-activity probability $\rho$ is small.

Notice further that by \eqref{eq:a}, the maximum URLLC DoF both in the inner and outer bounds is $\S^{(\U)}=\frac{\rho \rho_f}{2}$, because each Tx sends an URLLC message with probability $\rho \rho_f$ and in the deterministic setup of Section~\ref{sec:results}, the maximum URLLC DoF is $1/2$. Notice also that all bounds \eqref{eq:a}--\eqref{eq:conv2} increase 
 with the activity parameter $\rho$. The maximum eMBB DoF in the inner and outer bounds is $\S^{(\e)}= \rho - \frac{(1-\rho) \rho^{\D+2}}{1-\rho^{\D+2}}$. In the limit as $\D\to\infty$, it is thus given by $\rho$, and simply represents the expected fraction of active users. For finite $\D$, the eMBB DoF decreases because to avoid interference to propagate,  some of the Txs have to be silenced  as in the scheme of Subsection~\ref{sec:coding}. The term $\frac{(1-\rho) \rho^{\D+2}}{1-\rho^{\D+2}}$ thus describes the expected fraction of active but silenced Txs. 
 
\begin{figure}[t!]
\centering
\begin{subfigure}{0.45\textwidth}
\centering
\begin{tikzpicture}[scale=.8]
\begin{axis}[
    xlabel={\small {$\S^{(\U)}$ }},
    ylabel={\small {$\S^{(\e)}$ }},
     xlabel style={yshift=.5em},
     ylabel style={yshift=-1.25em},
    xmin=0, xmax=0.27,
    ymin=0, ymax=0.82,
    xtick={0,0.1,0.2,0.3,0.4},
    ytick={0,0.1,0.2,0.3,0.4,0.5,0.6,0.7,0.8,0.9,1},
    yticklabel style = {font=\small,xshift=0.25ex},
    xticklabel style = {font=\small,yshift=0.25ex},
    legend pos=south west,
]

%\addplot[
%    color=black,
%    mark=halfcircle,
%    ]
%    coordinates {
%    (0,0.9545)(0.0455, 0.9091)(0.5,0)
%    };%2
 \addplot[ color=green,   mark=star, line width = 0.5mm] coordinates {  (0,0.8) (0.24, 0.56)(0.24,0) };
 \addplot[ color=black,   mark=star, line width = 0.5mm, dashed] coordinates {  (0,0.7852) (0.24, 0.5452)(0.24,0) };
      \addplot[ color=blue,   mark=halfcircle, thick] coordinates {  (0,0.7852) (0.24, 0.5434)(0.24,0) };
   %   \addplot[ color=blue,   mark=diamond, thick, dashed] coordinates {  (0,0.7852) (0.12, 0.6627)(0.12,0) };
 %   \addplot[ color=black,   mark=halfcircle, thick] coordinates {  (0,0.7597) (0.24, 0.5131)(0.24,0) };
   \addplot[ color=brown,   mark=diamond,  line width = 0.5mm, dashed] coordinates {  (0,0.7289) (0.24, 0.4889)(0.24,0) };
     \addplot[ color=red,   mark=halfcircle, thick] coordinates {  (0,0.7289) (0.24, 0.4790)(0.24,0) };

%    \addplot[ color=black,   mark=diamond, thick, dashed] coordinates {  (0, 0.7597) (0.12, 0.6309)(0.12,0) };
  %   \addplot[ color=red,   mark=diamond, thick, dashed] coordinates {  (0,0.7289) (0.12, 0.5907)(0.12,0) };

%   \legend{{$ \rho_f = 0.6, \D = 10, M = 1.04$}, {$\rho_f = 0.6, \D = 6, M = -1.13$}, {$ \rho_f = 0.6, \D = 4, M = 1.28$}, {$ \rho_f = 0.3, \D = 10, M = 1.08$}, {$ \rho_f = 0.3, \D = 6, M = 1.29$}, {$ \rho_f = 0.3, \D = 4, M = 1.6$}}  
\small 
      \legend{{Outer and Inner Bounds, $\D = \infty$}, {Outer Bound, $\D = 10, M = 1$}, {Inner Bound, $ \D = 10, M = 1.006$},{Outer Bound, $\D = 4, M = 1$},  {Inner Bound, $\D = 4, M = 1.03$}}  
\end{axis}
%\draw [->,thick](2.3,4.7)--(2,3.65);
%\node [draw=none] (v1) at (2.2,5.4) {\small maximum-$\S^{(\e)}$};
%\node [draw=none] (v1) at (2.25,4.9) {\small boundary};
%\node [draw=none] (v1) at (5.3,2.3) {\small { maximum-$\S^{(\U)}$}};
%\node [draw=none] (v1) at (5.1,1.8) {\small {boundary}};
%\draw [->,thick](4,2.2)--(3,2.2);
%\node [draw=none, rotate = 90] (v1) at (0.5,2.2) {\small {\color{red}slope $= -3.86$}};
%\draw [->,thick, red](0.5,3.5)--(1.25,3.5);
%\node [draw=none, rotate = 0] (v1) at (4.3,3.9) {$\S^{(\e)} = \mu$};
\vspace{-0.4cm}
\end{tikzpicture}
\caption{$\rho_f = 0.6$.}
\label{fig2itw}
\end{subfigure}
\begin{subfigure}{0.45\textwidth}
\centering
\begin{tikzpicture}[scale=.8]
\begin{axis}[
    xlabel={\small {$\S^{(\U)}$ }},
    ylabel={\small {$\S^{(\e)}$ }},
     xlabel style={yshift=.5em},
     ylabel style={yshift=-1.25em},
    xmin=0, xmax=0.14,
    ymin=0, ymax=0.82,
    xtick={0,0.1,0.2,0.3,0.4},
    ytick={0,0.1,0.2,0.3,0.4,0.5,0.6,0.7,0.8,0.9,1},
    yticklabel style = {font=\small,xshift=0.25ex},
    xticklabel style = {font=\small,yshift=0.25ex},
    legend pos=south west,
]

%\addplot[
%    color=black,
%    mark=halfcircle,
%    ]
%    coordinates {
%    (0,0.9545)(0.0455, 0.9091)(0.5,0)
%    };%2
 \addplot[ color=green,   mark=star, line width = 0.5mm] coordinates {  (0,0.8) (0.12, 0.68)(0.12,0) };
 \addplot[ color=black,   mark=star, line width = 0.5mm, dashed] coordinates {  (0,0.7852) (0.12, 0.6652)(0.12,0) };
 %     \addplot[ color=blue,   mark=halfcircle, thick] coordinates {  (0,0.7852) (0.24, 0.5432)(0.24,0) };
      \addplot[ color=blue,   mark=halfcircle, thick] coordinates {  (0,0.7852) (0.12, 0.6631)(0.12,0) };
 %   \addplot[ color=black,   mark=halfcircle, thick] coordinates {  (0,0.7597) (0.24, 0.5131)(0.24,0) };
    % \addplot[ color=red,   mark=halfcircle, thick] coordinates {  (0,0.7289) (0.24, 0.4750)(0.24,0) };

 \addplot[ color=brown,   mark=diamond,  line width = 0.5mm, dashed] coordinates {  (0,0.7289) (0.12, 0.6089)(0.12,0) };
     \addplot[ color=red,   mark=halfcircle, thick] coordinates {  (0,0.7289) (0.12, 0.5965)(0.12,0) };

%   \legend{{$ \rho_f = 0.6, \D = 10, M = 1.04$}, {$\rho_f = 0.6, \D = 6, M = -1.13$}, {$ \rho_f = 0.6, \D = 4, M = 1.28$}, {$ \rho_f = 0.3, \D = 10, M = 1.08$}, {$ \rho_f = 0.3, \D = 6, M = 1.29$}, {$ \rho_f = 0.3, \D = 4, M = 1.6$}}  
\small 
      \legend{{Outer and Inner Bounds, $\D = \infty$}, {Outer Bound $\D = 10, M = 1$},{Inner Bound, $\D = 10, M = 1.01$}, {Outer Bound $\D = 4, M = 1$}, {Inner Bound, $\D = 4, M = 1.08$}}  
\end{axis}

%\draw [->,thick](2.3,4.7)--(2,3.65);
%\node [draw=none] (v1) at (2.2,5.4) {\small maximum-$\S^{(\e)}$};
%\node [draw=none] (v1) at (2.25,4.9) {\small boundary};
%\node [draw=none] (v1) at (5.3,2.3) {\small { maximum-$\S^{(\U)}$}};
%\node [draw=none] (v1) at (5.1,1.8) {\small {boundary}};
%\draw [->,thick](4,2.2)--(3,2.2);
%\node [draw=none, rotate = 90] (v1) at (0.5,2.2) {\small {\color{red}slope $= -3.86$}};
%\draw [->,thick, red](0.5,3.5)--(1.25,3.5);
%\node [draw=none, rotate = 0] (v1) at (4.3,3.9) {$\S^{(\e)} = \mu$};

\vspace{-0.4cm}
\end{tikzpicture}

\caption{$\rho_f = 0.3$.}
\label{fig3itw}
\end{subfigure}
\caption{Inner and outer bounds on  the DoF region  for $\rho= 0.8$ and different values of $\D$.}
\label{figitw}
\end{figure}
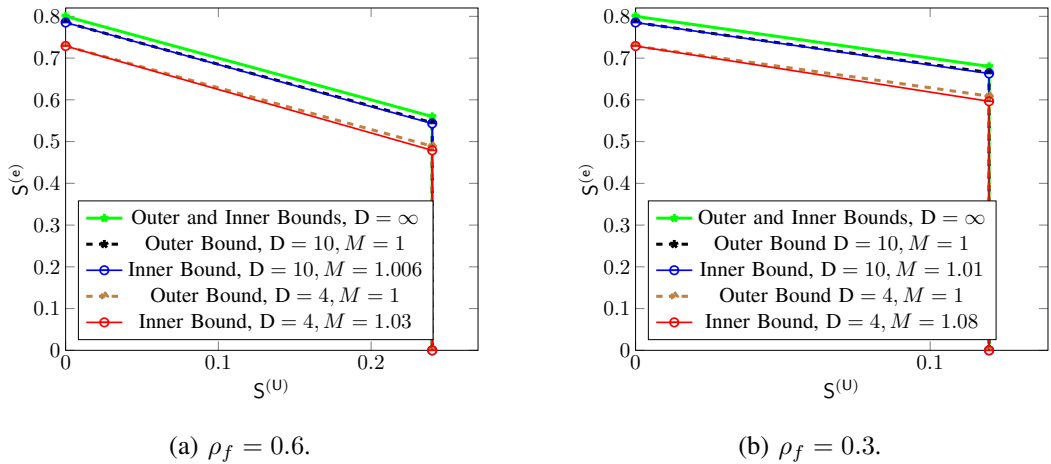

 Figures \ref{figitw},  illustrate the inner and outer bounds in  \eqref{eq:a}--\eqref{eq:conv2}  for different values of $\rho, \rho_f$, and $\D$. 
 The  most interesting part of the plots is the upper side of the trapezoids (the side lying opposite the two right angles). The slope of this line, which is $-1$ for the outer bounds and $-M$ for the inner bounds,  describes the penalty in maximum eMBB DoF $\S^{(\e)}$ incurred when one  increases the URLLC Dof $\S^{(\U)}$. Thus, on the outer bounds, increasing $\S^{(\U)}$ by $\Delta$ decreases the maximum $\S^{(\e)}$  by $\Delta$ and thus the sum DoF stays constant for all values  of $\S^{(\U)}$. On the inner bounds, the maximum eMBB DoF $\S^{(\e)}$ is decreased by $M \Delta>\Delta$ when $\S^{(\U)}$ is increased by $\Delta$ and the sum DoF thus decreases by  $(M-1) \Delta>0$. % As seen on the figures 
%Figures~\ref{fig2itw} and \ref{fig3itw} further indicate that the penalty in maximum sum-MG of our inner bounds also decreases when the ``fast" activity parameter $\rho_f$ increases. For example,  for $\rho=0.6$ and $\D = 4$  the sum-MG penalty $(M-1)$ of the inner bound decreases from $0.08$ for $\rho_f=0.3$ to $0.03$ for $\rho_f=0.6$ (see Figures \ref{fig3itw} and \ref{fig2itw}).%By equation \eqref{eq:M}, the slope increases as $\rho_f$ increases. Note that, the term $\Delta$ decreases as $\rho_f$ increases but for a given pair $(\rho, \rho_f)$ each in $(0,1)$ and $\D \ge 2$, $\Delta < \rho \rho_f (1- \rho^2 (1- \rho_f))$ and thus in general the slope increases with $\rho_f$. 

%-------------------------------

%Txs and Rxs in the \emph{inactive set} $\mathcal{K}\backslash \Ta$ do not participate in the cooperation phases. Notice that all our results remain valid in a setup where inactive Txs and Rxs \emph{do} participate in the cooperation phases. Since our inner and outer bounds are rather close in general (see the subsequent numerical discussion), this indicates that without essential loss in optimality Txs and Rxs in $\mathcal{K}\backslash \Ta$  can entirely be set to sleep mode to conserve their batteries.
 
A similar model, but with only Rx-cooperation was studied in \cite{HomaITW2020} for Wyner's soft-handoff model, Wyner's linear symmetric model, and the hexagonal model.  Similarly to the setup with deterministic user activities and arrivals, with only Rx-cooperation but no Tx-cooperation the stringent delay requirement on URLLC messages harm the overall performance of the system and maximum sum DoF is only achieved when  transmitting only URLLC messages.  For networks with regular interference structures as in Figure~\ref{fig5.10} and when $\rho \rho_f \gg 1$, e.g., because URLLC messages are rare, it was shown in \cite{HomaITW2020} that DoF pairs $(\S^{(\U)}, \S^{(\e)})$ satisfying the following equality are achievable
\begin{equation}
\S^{(\e)} = \rho - (1 + \ell \rho) \S^{(\U)},
\end{equation} 
for $\S^{(\U)} \leq \frac{\rho \rho_f}{\ell}$ and $\ell$ denoting the number of interference signals experienced at each Rx. (For example, for Wyner's linear symmetric network $\ell=2$ and for the hexagonal model $\ell=6$.)

\subsection{Summary}\label{sec:Coop_summary}
This section  considered cooperative interference networks where only eMBB messages can profit from these cooperation links, but not  URLLC messages because they have to be transmitted and decoded without further due.  A general coding scheme was presented that manages to exploit the cooperation links for the transmission of eMBB messages in a way that  allows to attain  the optimal overall performance (sum DoF) of the system despite the transmission of URLLC messages. Achieving this performance requires cooperation both at the Tx and Rx side, one of the two is not sufficient. Moreover,  a careful scheduling of  URLLC and eMBB messages to users had to be performed. In practice this scheduling is performed at a system level in the sense that applications randomly generate URLLC messages. The proposed scheme in \cite{HomaITW2021} was adapted to such random and bursty arrivals of the URLLC messages, with only a small penalty in overall system performance. %The situation is however different when only transmitters or only receivers can cooperate, where information-theoretic converse argument establish even integrated coding schemes suffer from a penalty in the overall-system performance when cooperation is only one-sided. 
%------------------------------

\section{Cloud Radio Access Networks (C-RANs)}\label{sec:CRAN}

%\section
\subsection{Introduction} 
%\begin{figure}[t!]
%\begin{center}
%\includegraphics[width=0.5\textwidth]{figures/CRAN2.pdf}
%\end{center}
%\caption{Cloud radio access networks}
%\label{fig:CRAN2}
%\end{figure}
In this section we consider cloud radio access networks (C-RANs) where BSs are connected to a cloud processor via high-rate fronthaul links and in the uplink communication all transmitted messages are jointly decoded  at the central processor so as to be able to alleviate the effect of interference \cite{Peng2015}.   URLLC messages are however not compatible with this new architecture as they have to be decoded directly at  the BSs because communication over an additional hop to the cloud processor would violate their stringent delay constraints. As in the previous sections, we wish to investigate how this restriction affects the overall performance of the systems, and more specifically the sum-rates and rate pairs can simultaneously be achieved for URLLC and eMBB transmissions. 
 
 This section reviews two pieces of work on C-RAN with mixed-delay traffic. 
%Both model the communication from mobiles to BSs by Wyner's softhandoff model. 
The work in \cite{HomaITW2019}, discussed in Subsection \ref{sec:CRAN1} considers a fading network and focuses on an information-theoretic discussion of the problem comparing inner and outer bounds on the fundamental performance limits of such systems. The work in \cite{Kassab2018, Kassab2019}, discussed in Subsection \ref{sec:CRAN2}, takes a communication-theoretic approach. It decomposes communication in minislots and then compares performance of  different communication strategies. In this latter work, the network is assumed static. 
%Consider a C-RAN system where each mobile in each cell has either an URLLC message and/or an eMBB message to decode. Transmitted URLCC messages are directly decoded at the BSs. 
%EMBB messages are jointly decoded at the cloud processor, where to this end, BSs compress their output signals and send the compression indices to the cloud processor. 

\subsection{Fading C-RAN Model}\label{sec:CRAN1}
The work in \cite{HomaITW2019}  considers the uplink of a C-RAN, and models the network from the  mobile users to the BSs by an i.i.d. fading Wyner soft-handoff model, see Figure~\ref{fig:CRAN_Wyner}. That means, BSs are aligned on a line and each cell contains  a single mobile user. This latter assumption stems again from the orthogonal frequency access applied by various mobile users in a cell. In Wyner's soft-handoff model, mobile users are  assumed to be located on cell borders and thus interfere only on the communication in this neighbouring and closeby cell.
At a given time $t\in[n]$, the signal received at any BS~$k\in[\K]$ is  thus described as
\begin{equation}\label{Eqn:Channel}
Y_{k,t} =G_{k,t} X_{k,t} + F_{k,t} X_{k-1,t} +Z_{k,t},
\end{equation}
where $X_{k,t}$ and $X_{k-1,t}$ are the signals sent by mobile users $k$ and  $k-1$ at time $t$; $\{Z_{k,t}\}$ are  i.i.d  standard Gaussian noise; and 
 the sequence of   channel coefficients 
\begin{equation}
\big \{(G_{1,t}, G_{2,t},\ldots, G_{\K,t},F_{1,t}, F_{2,t},\ldots, F_{\K,t})\big \}_{t=1}^{n}
\end{equation} is i.i.d. over time and distributed according to a given $\K$-tuple distribution $P_{G_1\cdots G_{\K}F_1\cdots F_{\K}}$. 
This $\K$-tuple distribution is the marginal distribution of a given stationary and ergodic process $\{(G_k,F_k)\}_{k=-\infty}^{\infty}$
satisfying $\E{ |G_0|^2 }< \infty$ and $\E {|F_0|^2} < \infty$. 
%\begin{align}\label{eq:assumptions}
%&\E{ |G_0|^2 }< \infty, \quad \quad \quad \quad \quad \E {|F_0|^2} < \infty, \nonumber \\
%&\E{\frac{ |G_0|^2}{1+|F_0|^2}} < \infty \quad \text{and} \quad \E{\frac{ |F_0|^2}{1+|F_0|^2}}< \infty.
%\end{align}
%where $||.||$ is the spectral norm.
 % across users. 
%Also, denote the  $P_{GF}$ single
%\begin{equation} 
%(G_{k,t},F_{k,t}) \sim P_{GF}, \qquad \forall k\in\{1,\ldots, K\}, \ t\in\{1,\ldots,n\}.
%\end{equation}
The fading coefficients  $\{(G_{k,t}, F_{k,t})\}$ are known perfectly at BS $k$ but not at the mobile users.

 Each mobile user $k$ sends both an URLLC message $M_k^{(\U)}$ and an eMBB message $M_k^{(\e)}$.   It thus produces its  channel inputs  as $\mathbf{X}_k=f_k\big(M_k^{(\U)},M_k^{(\e)}\big)$, and so that they satisfy an average block-power constraint $\P$.  URLLC messages are directly decoded at the BSs  based on the observed signals in \eqref{Eqn:Channel}.  eMBB messages are decoded at the cloud processor, which perfectly observes the symbols sent over the fronthaul links by the  BSs, where each BS~$k$ generates its symbols $\vect{L}_k$ by employing a compression function $f_k$ to its observed outputs $Y_k^n$. The compression has to account for the capacity of the fronthaul links, which is assumed to be  $C=\mu \frac{1}{2} \log (1+\P)$ for each link, where   $\mu$ is termed the fronthaul prelog.
\begin{figure}[t]
%\small
  \centering
  \small
    %\hspace*{32pt}
 \begin{tikzpicture}[scale=1.47, >=stealth]
\centering
\tikzstyle{every node}=[draw,shape=circle, node distance=1cm];
\draw (-0.5,0) -- (0.5,0) -- (0.5,0.5) -- (-0.5,0.5) -- (-0.5,0);
\draw (-0.5,2) -- (0.5,2) -- (0.5,2.5) -- (-0.5,2.5) -- (-0.5,2);
\draw (1.5,2) -- (2.5,2) -- (2.5,2.5) -- (1.5,2.5) -- (1.5,2);
\draw (1.5,0) -- (2.5,0) -- (2.5,0.5) -- (1.5,0.5) -- (1.5,0);
\draw (3.5,2) -- (4.5,2) -- (4.5,2.5) -- (3.5,2.5) -- (3.5,2);
\draw (3.5,0) -- (4.5,0) -- (4.5,0.5) -- (3.5,0.5) -- (3.5,0);
\node [draw] at (0,1.25) {$+$};
\node [draw] at (2,1.25) {$+$};
\node [draw] at (4,1.25) {$+$};
\draw   [thick,<-] (0,2)--(0,1.49);
\draw   [thick,<-] (2,2)--(2,1.49);
\draw   [thick,<-] (4,2)--(4,1.49);
\draw   [thick,<-] (0,1.02)--(0,0.5);
\draw   [thick,<-] (2,1.02)--(2,0.5);
\draw   [thick,<-] (4,1.02)--(4,0.5);
\draw   [thick,->] (0.75,1.25)--(0.23,1.25);
\draw   [thick,->] (2.75,1.25)--(2.23,1.25);
\draw   [thick,->] (4.75,1.25)--(4.23,1.25);
\draw   [thick,->,dashed] (0,0.5)--node [draw=none,  text width=3.0cm, shape=rectangle, pos=0.5, yshift=0cm,xshift=1cm ,rotate = 20,midway, fill=none, node distance=1cm] {\small{$F_k X_{k-1}^n$}} (1.76,1.25);
\draw   [thick,->,dashed] (2,0.5)--node [draw=none,  text width=3.0cm, shape=rectangle, pos=0.5, yshift=0cm,xshift=0.9cm ,rotate = 20,midway, fill=none, node distance=1cm] {\small{$F_{k +1} X_{k}^n$}}(3.76,1.25);
\draw   [thick,->,dashed] (-1,0.8)--(-0.19,1.25);
%%%%%%%
\node [draw=none] at (-0.3,1.75) {\small{$Y_{k-1}^n$}};
\node [draw=none] at (1.8,1.75) {\small$Y_{k}^n$};
\node [draw=none] at (3.73,1.75) {\small$Y_{k+1}^n$};
\node [draw=none] at (0.5,1.39) {\small$Z_{k-1}^n$};
\node [draw=none] at (2.5,1.39) {\small$Z_{k}^n$};
\node [draw=none] at (4.5,1.39) {\small$Z_{k+1}^n$};
\node [draw=none] at (-0.3,0.75) {\small$X_{k-1}^n$};
\node [draw=none] at (1.8,0.75) {\small$X_{k}^n$};
\node [draw=none] at (3.73,0.75) {\small$X_{k+1}^n$};
\node [draw=none] at (0,0.25) {\small MU $k-1$};
\node [draw=none] at (2,0.25) {\small MU $k$};
\node [draw=none] at (4,0.25) {\small MU $k+1$};
\node [draw=none] at (0,2.25) {\small BS $k-1$};
\node [draw=none] at (2,2.25) {\small BS $k$};
\node [draw=none] at (4,2.25) {\small BS $k+1$};
\draw [very thick, blue, <-](2,3.35)--node [draw=none,  text width=0.5cm, shape=rectangle, pos=0.5, yshift=-0.2cm,xshift=-0.2cm ,rotate = 0,midway, fill=none, node distance=1cm] {\small{$L_k $}}(2,2.5);
\draw [very thick, blue, <-](2-0.8,4-0.5)--node [draw=none,  text width=1.4cm, shape=rectangle, pos=0.5, yshift=-0.2cm,xshift=-0.45cm ,rotate = 0,midway, fill=none, node distance=1cm] {\small{$L_{k-1} $}}(0,2.5);
\draw [very thick, blue, <-](2+0.85,4-0.55)--node [draw=none,  text width=-1cm, shape=rectangle, pos=0.5, yshift=-0.2cm,xshift=0cm ,rotate = 0,midway, fill=none, node distance=1cm] {\small{$L_{k+1} $}}(4,2.5);
\draw [->, blue, very thick] (2, 4.5)--(2,5.25);
\draw [->, blue, very thick] (1.25, 4.5)--(1.25,5.25);
\draw [->, blue, very thick] (2.75, 4.5)--(2.75,5.25);
\node[cloud, cloud puffs=15.7, cloud ignores aspect, minimum width=5cm, minimum height=2cm, align=center, draw, fill= blue!30] (cloud) at (2cm, 4cm) {CU};
\node [draw=none] at (0.8,2.45) {{\color{red}\small $M_ {k-1}^{(\U)}$}};
\draw [->, red, very thick] (0.5, 2.25)--(1.1,2.25);
\node [draw=none] at (0.8+2,2.45) {{\color{red}\small $M_ {k}^{(\U)}$}};
\draw [->, red , very thick] (0.5+2, 2.25)--(1.1+2,2.25);
\node [draw=none] at (0.8+4,2.45) {{\color{red}\small $M_ {k+1}^{(\U)}$}};
\draw [->, red, very thick] (0.5+4, 2.25)--(1.1+4,2.25);
\node [draw=none] at (1.25,5.4) {{\color{blue}\small $M_ {k-1}^{(\e)}$}};
\node [draw=none] at (2,5.4) {{\color{blue}\small $M_ {k}^{(\e)}$}};
\node [draw=none] at (2.75,5.4) {{\color{blue}\small $M_ {k+1}^{(\e)}$}};
\node [draw=none] at (3.25,5) {{\color{blue}$\ldots$}};
\node [draw=none] at (0.75,5) {{\color{blue}$\ldots$}};
%\node [draw=none] at (2,5) {\small{Cloud}};

\end{tikzpicture}
%\vspace*{-5ex}

  \caption{C-RAN with URLLC and eMBB transmissions and  the mobile-to-BS network modeled by Wyner's soft-handoff model. }
  \label{fig:CRAN_Wyner}
  \vspace*{-2ex}
\end{figure}
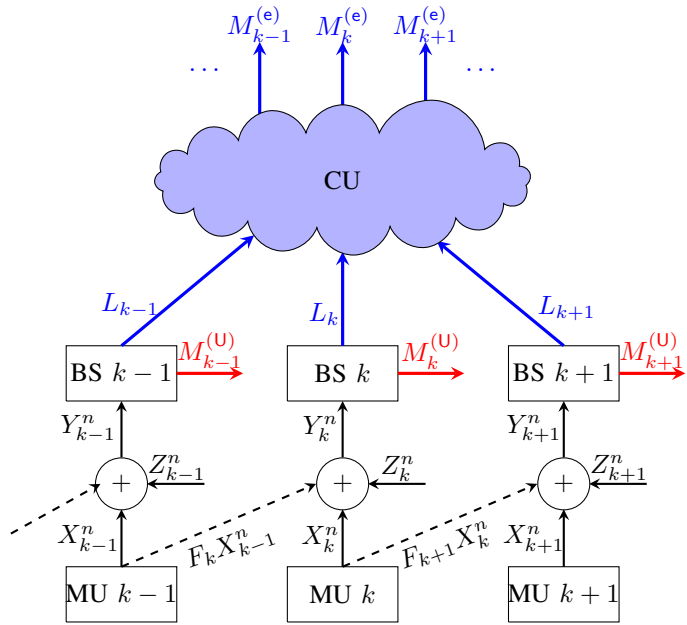~~%

For the described model,  \cite{HomaITW2019} characterizes the set  of all achievable  average expected  URLLC and eMBB DoFs (across users and fadings) as
 \begin{subequations}\label{eq:DoFCRAN}
 \begin{IEEEeqnarray}{rCl}
 2 \S^{(\U)} + \S^{(\e)} \leq 1\\
 \S^{(\e)} \leq \mu.
 \end{IEEEeqnarray}
 \end{subequations}
The entire DoF region can be achieved by a resource scheduling approach that time-shares URLLC and eMBB transmissions. Notice that the eMBB DoF is limited by the fronthaul  prelog $\mu$ because  all eMBB messages have to be decoded at the cloud center. For small fronthaul prelogs $\mu$ this restriction limits the DoF of the system, which can be improved by allowing BSs to also decode part of the eMBB messages. 

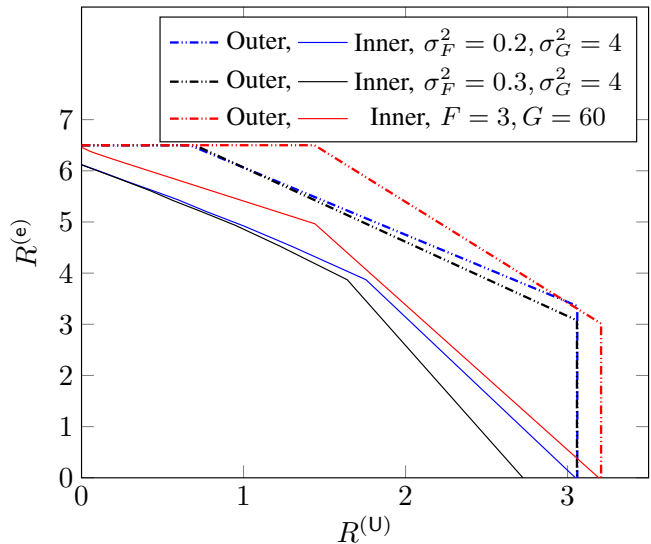
\begin{figure}[!t]
%%\psfragscanon
\centering
\begin{tikzpicture}[scale=1.1]
\begin{axis}[
    xlabel={\small {$R^{(\U)}$ }},
    ylabel={\small {$R^{(\e)}$ }},
     xlabel style={yshift=.5em},
     ylabel style={yshift=-1.25em},
    legend columns=2,
    xmin=0, xmax=3.5,
    ymin=0, ymax=9.2,
   xtick={0,1,2,3,4, 5,6},
    ytick={0,1,2,3,4, 5,6,7},
    yticklabel style = {font=\small,xshift=0.25ex},
    xticklabel style = {font=\small,yshift=0.25ex},
]

\addplot[color=blue, densely dash dot dot, thick] coordinates {(0,6.5000)    (0.6685,6.5000)    (3.0599,3.3570)    (3.0599,0)
 };%1 
\addplot[ color=blue] coordinates {(0,6.1198)    (0.0927,6.0146)    (0.1951,5.8929)   (0.3101,5.7581)    (0.4407,5.6117)    (0.5928,5.4366)    (0.7715,5.2019)    (0.9977,4.9262)    (1.2965,4.5255)    (1.7556,3.8713)    (3.0485,0)};%1  

\addplot[color=black, densely dash dot dot, thick] coordinates {(0,6.5000)    (0.6998,6.5000)    (3.0579,3.0754)    (3.0579,0)
};
 \addplot[ color=black] coordinates {(0,6.1170)    (0.0891,6.0112)    (0.1877,5.8954)    (0.2975,5.7636)    (0.4222,5.6131)    (0.5651,5.4183)    (0.7364,5.2067)    (0.9494,4.9339)    (1.2241,4.5278)    (1.6403,3.8699)    (2.7245,0)};%1  

  \addplot[color=red, densely dash dot dot, thick] coordinates {(0,6.5000)    (1.4383,6.5000)   (3.2067,3.0147)    (3.2067,0)};  
    \addplot[ color=red] coordinates { (0,6.4618)    (0.0525,6.3805)    (0.1113,6.3145)    (0.1778,6.2442)    (0.2546,6.1650)    (0.3453,6.0727)    (0.4563,5.9598)    (0.5991,5.8150)    (0.7998,5.6115)   (1.1404+0.3,5.2643-0.3)    (3.1926,0)
};%1   

\legend{{\footnotesize Outer,}, {\footnotesize Inner, $\sigma_F^2 = 0.2, \sigma_G^2 = 4$}, {\footnotesize Outer, }, {\footnotesize Inner, $\sigma_F^2 = 0.3, \sigma_G^2 = 4$},{\footnotesize Outer, }, {\footnotesize Inner, $F = 3, G = 60$}};
\end{axis}
\end{tikzpicture}
\caption{Inner and outer bounds on the achievable $R^{(\U)}$ and $R^{(\e)}$ rate region presented in \cite{HomaITW2019} for $\P=100$, $\C = 6.5$, Gaussian i.i.d. fadings of variances $\sigma_F^2$ and $\sigma_G^2$ or for constant non-time varying fadings  $F=3$ and $G=60$.}
\label{fig:CRAN_plot}
\end{figure}

As can be inferred  from \eqref{eq:DoFCRAN}, the DoF region of the described C-RAN does not depend on the fading processes $\{F_{k,t}\}$ and $\{G_{k,t}\}$. This contrasts the behaviour at finite powers $\P$ where the set of achievable average URLLC and eMBB rates depends on the law of this fading process, as can be seen in Figure~\ref{fig:CRAN_plot}. The performance described in this Figure~\ref{fig:CRAN_plot} is attained by a superposition coding scheme, and significantly improves over a pure scheduling scheme.
We further notice from the figure that for small values of URLLC rates  $R^{(\U)}$, when $R^{(\U)}$ increases by $\Delta$, then the maximum eMBB rate $R^{(\e)}$ decreases approximately by the same amount $\Delta$, and thus the sum-rate remains constant. For larger values of $R^{(F)}$, the maximum $R^{(\e)}$ decreases approximately by $3\Delta$ if $R^{(\U)}$ is increased by $\Delta$.  Thereby the loss is larger for random than for static fading coefficients.

\subsection{Static CRAN Model with Slotted Communication}\label{sec:CRAN2}

Mixed delay constraints in C-RANs were also studied in \cite{Kassab2018}, where on a system-wide level communication is divided into minislots. In each  minislot, each mobile user generates an URLLC message with probability $q$ and attempts to send it over the network during the next minislot that is dedicated to URLLC communication. If a user generates multiple URLLC messages before the next URLLC minislot, it drops all but one URLLC message, which is then sent in this minislot. eMBB messages are sent over multiple minislots and share the available resources in eMBB slots. Moreover, in \cite{Kassab2018}, URLLC communication is assumed to be from mobile users close to the BSs, and their communication  does not suffer from intercell-interference. eMBB users are assumed on the network border as in \cite{HomaITW2019} and communication suffers from intercell interference as described by Wyner's symmetric model.

\begin{figure}[t!]
\vspace{-2cm}
\begin{center}
\includegraphics[width=0.5\textwidth]{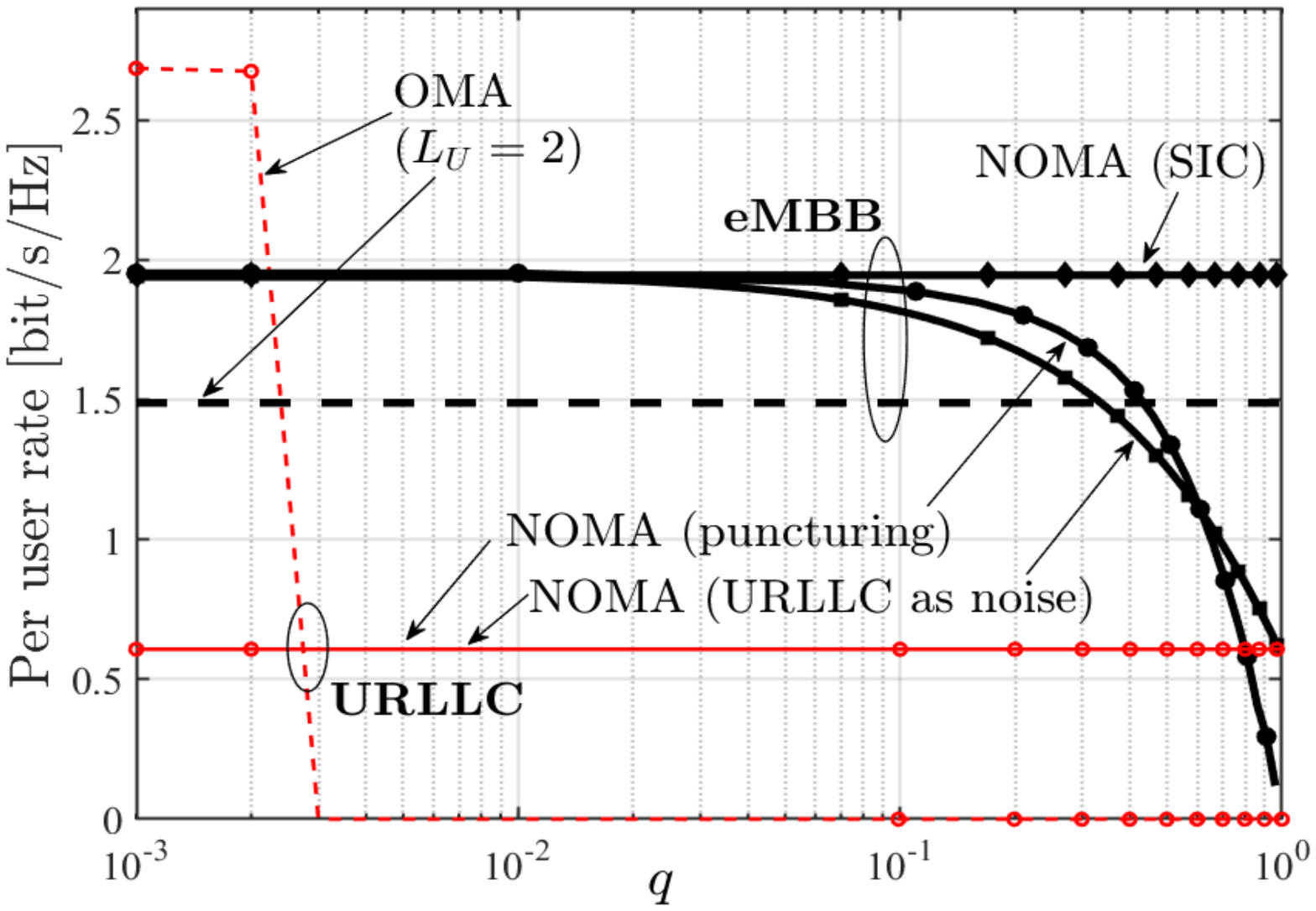}
\end{center}
\vspace{-3cm}
\caption{eMBB and URLLC per-user rates under OMA with
$L_{\U}$ and NOMA for different decoding strategies as
function of $q$ in C-RAN \cite{Kassab2018}.}
\label{fig:CRANplot_2}
\end{figure}

The performances of different coding schemes are compared in \cite{Kassab2018}. In all the schemes, eMBB messages are transmitted   using standard multi-access codes, since  they are jointly decoded at the cloud processor. URLLC messages are transmitted using standard Gaussian codebooks.  If eMBB and URLLC messages are sent in the same slots, then eMBB communication (which lasts several minislots) is considered as noise in the decoding of URLLC messages. 
\begin{itemize}
\item \textit{OMA:} URLLC and eMBB messages are sent using \emph{orthogonal multiaccess (OMA)}, i.e., a pure scheduling approach where the minislots dedicated for URLLC and eMBB communications are disjoint. Specifically, here every $L_{\U}$-th minislot is dedicated for URLLC transmission and the other minislots for eMBB transmissions.
\item \textit{NOMA--puncturing:} eMBB and URLLC messages are sent using \emph{non-orthogonal multiaccess (NOMA)}, i.e.,  eMBB  and URLLC are sent over the same minislots. In particular, URLLC messages are transmitted in the minislot following their generation. To avoid this URLLC communication interfering with eMBB communication, BSs compress   only  the signals they receive  in minislots where no URLLC communication is  taking place from their corresponding mobile user. 
\item \textit{NOMA--treating URLLC as noise:} eMBB and URLLC messages are sent using NOMA. URLLC communication is treated as noise for eMBB  decoding. Therefore, BSs compress all their output signals, and send all this compression information to the cloud processor. 
\item \textit{NOMA--SIC:} As in the previous item, except that BSs perform \emph{successive interference cancellation (SIC)} on their decoded URLLC messages. That means, if URLLC decoding is successful, they subtract the URLLC signal from their outputs before compressing it for transmission to the cloud processor. 
\end{itemize}
The performance of eMBB transmissions is measured by the information-theoretic rate that is achievable  in the asymptotic regime of large blocklengths.
 URLLC transmissions are performed over single minislots and  thus of much smaller blocklength. Their performances are measured using a finite blocklength rate-expression \cite{Yuri2012}  
\begin{equation}
R_{\U} = \log (1+ S_{\U}) - \sqrt{\frac{S_{\U}}{n(1+S_{\U})} } \mathcal{Q}^{-1}\left( \epsilon_{\U}\right),
\end{equation}
where $S_{\U}$ denotes the interference power at a BS and $\epsilon_{\U}$ has to be chosen sufficiently small so that the overall error probability (including the packet drops in case of OMA) does not exceed a desired threshold. Figure~\ref{fig:CRANplot_2}  compares the performances of these schemes in function of the URLLC message generation probability $q$. We observe that for the OMA approach  performance degenerates quickly even for $L_{\U}=2$ because the probability of URLLC message drop  is too large. The URLLC performance is identical for all three NOMA approaches. The NOMA SIC approach achieves the best performance for the eMBB message. 

In \cite{Kassab2019}, these results are also extended to the downlink scenario. In this case, eMBB messages are created at the cloud processor and can profit from joint encoding to mitigate interference.  URLLC messages are created directly at the BSs and their communications thus suffer from interference.

\subsection{Summary}\label{sec:CRAN_summary}
The last setup considered in this paper are C-RAN architectures where eMBB messages are jointly decoded at the cloud processor, whereas  URLLC messages have to be decoded immediately   at the BSs.  Similarly to the P2P, BC, and cooperative network scenarios, for moderate powers, the overall  system performance  of the C-RAN with mixed-delay traffic in Subsection~\ref{sec:CRAN1} decreases for large URLLC rates. This degradation seems to be  more pronounced in fading environments than in static Gaussian environments. In the asymptotic high-power regime, however such a degradation is not observed, and at small URLLC DoFs $\S^{(\U)}$, the sum-DoF is even increasing in  $\S^{(\U)}$.  In certain scenarios it is thus possible to improve overall system performance by decoding part of the eMBB messages directly at the BSs and not at the cloud processor.  Subsection~\ref{sec:CRAN2} considers a non-fading environment and random generation of URLLC messages, for which it applies finite-blocklength performance measures. It is shown that for moderate or high URLLC generation rates, a NOMA scheme that first subtracts the contribution of the URLLC communication from the receive signals at the BSs, and then compresses  and sends these differences over the fronhaul links, outperforms similar OMA and NOMA schemes.

\section{Conclusions and Outlook}\label{sec:Summary}
%%Summary
In this survey, we have reviewed joint  coding schemes that integrate transmissions of  URLLC and eMBB traffic and compared them to pure scheduling schemes in terms of rate, error probability and degree of freedom  pairs that the schemes simultaneously achieve for  URLLC and eMBB messages. A wide range of communication scenarios including P2P channels, BCs,   cooperative interference  networks, and C-RANs have been considered. The results have shown that  joint coding schemes can significantly outperform the standard scheduling approach. As we have seen, in. certain scenarios optimal system performance can be achieved under any URLLC rate. For other scenarios however, a large URLLC rate penalizes the overall system performance, showing that in these situations the stringent URLLC decoding constraint degrades  system performance.

We conclude this survey with some lines of  potential future research. 
\begin{itemize}
 \item  The presented works have considered perfect  channel state information (CSI) at the Rxs, and sometimes even at the Txs, where naturally CSI is more difficult to obtain.  An interesting model for mixed-delay traffic is where CSI can be used for encoding and decoding of eMBB messages but not of URLLC messages \cite{Wangetal, Chan2019}. The motivation behind such a model is that the  processing of pilot  and feedback signals required to gather CSI at the Rxs and Txs introduces inadmissibly  large delays for URLLC communication.

\item So far, firm finite-blocklength results for mixed-delay traffic have been mostly limited to the P2P case; see   \cite{Mary2016} for an exception. Extensions to multi-user network scenarios is an important future research direction.

\item In practical scenarios, both URLLC and eMBB messages are randomly generated by higher layer applications. This naturally leads to  potential bottlenecks where not-yet-transmitted messages have to be buffered, similar to\cite{Steiner2010}. In this context, a thorough analysis of the behavior of the buffered contents and the required  size of these buffers, is of high practical interest.

\item Other heterogeneous requirements on URLLC and eMBB traffic could be introduced in the  study of mixed-delay traffic For example, different security requirements as in \cite{Zou2018} or different reliability constraints as in \cite{Keresztfalvi2019}. 

\mw{
\item Langberg and Effros \cite{Langberg2021} introduced the notion of \emph{time-rate region} that describes the  fraction of the blocklengths required for the transmissions of the various messages to the different Rxs in a network scenario under given message communication rates.  A natural question is whether the interference mitigation techniques discussed in this survey can improve the inner bound on the time-rate region for general networks  obtained in \cite{Langberg2021}, which is obtained through a reduction to standard network information theory problems.

\item Finally, mixed-delay traffic where different messages are transmitted over different blocklengths is inherently also connected to variable-rate and variable-length coding \cite{Verdu2010,Shulman2000,Langberg2021,Steinberg2017}. For example, the variable-rate channel coding framework of \cite{Verdu2010} includes the variable-to-variable scenario where depending on the specific system configuration and channel state realization, a receiver can decode a message of variable-size (similarly to the broadcast approaches in Sections III-B and IV) and decoding is performed after a variable number of channel uses. An interesting line of future research is to extend this scenario to multiple messages and mixed-delay traffic where URLLC and eMBB messages are decoded with different delays, and to study the four-dimensional tradeoff between URLLC and eMBB variable-rates and variable-delays.}

%\item In the completely centralized architecture of C-RAN, we introduced a setup where neither BSs nor UEs can cooperate and we only focused on Wyner's soft-handoff model. Our next steps could be to: (i) investigate the influence of the model employed for the network between UEs and BSs, as well as the availability of cooperation links between UEs and/or BSs; and (ii) study mixed delay constraints on transmissions over fog radio access networks (FRAN) where large amount of signal processing and computing is also performed in BSs and UEs. Readers are encouraged to refer to \cite{Karasik2019}. 
%
%In the scheme proposed for C-RAN, we used simple point-to-point compression techniques. One can  
%reduce the gap between the proposed inner and outer capacity bounds by including more sophisticated multi-user compression techniques such as successive Wyner-Ziv compression and noisy network coding \cite{Aquerri2019,osvaldo} at the BSs and the cloud processor.
\end{itemize}

\section*{Acknowledgment}
The works of H. Nikbakht, M.~Egan and J-M.~Gorce have been supported by INRIA Nokia Bell Labs ADR “Network Information Theory" 
and  by the French National Agency for Research (ANR) under grant ANR-16-CE25-0001 - ARBURST.
The work of M. Wigger has been supported by the
       European Union's Horizon 2020 Research And Innovation Program,
       grant agreements no. 715111. The work of S. Shamai (Shitz) has been supported
by the US-Israel Binational Science Foundation
(BSF) under grant BSF-2018710. The work of H. V. Poor
has been supported by the U.S. National Science Foundation (NSF) within the Israel-US Binational program
under grant CCF-1908308.
%	\bibliographystyle{ieeetr}
%	
%	\bibliography{homa_tutorial.bib}

\end{document}